\documentclass{article}
\usepackage{authblk}
\usepackage{graphicx} 
\usepackage{color}
\usepackage{geometry}
\usepackage{lmodern}
\usepackage[size=tiny]{todonotes}
\usepackage{multirow}
\usepackage{float}
\usepackage{tabularx}
\usepackage{adjustbox}
\usepackage{subcaption}
\usepackage{xcolor}
\usepackage{hyperref}
\usepackage{comment}
\usepackage{array}
\usepackage{amsmath,amssymb}
\usepackage{csvsimple} 
\usepackage{hyperref}
\usepackage{xcolor}
\usepackage{longtable} 
\usepackage{arydshln}
\usepackage{xcolor}
\usepackage{tikz} 
\usepackage{booktabs} 
\usetikzlibrary{calc, arrows.meta, intersections, patterns, positioning, shapes.misc, fadings, through,decorations.pathreplacing}

\definecolor{ColorOne}{named}{blue}
\definecolor{ColorTwo}{named}{orange}
\definecolor{ColorThree}{named}{purple}

\makeatletter

\define@key{todonotes}{zr}[]{%
    \setkeys{todonotes}{author=zr,color=green}}%
\define@key{todonotes}{ak}[]{%
    \setkeys{todonotes}{author=ak,color=pink}}%

\makeatother

\newcolumntype{P}[1]{>{\centering\arraybackslash}p{#1}}
\newcolumntype{M}[1]{>{\centering\arraybackslash}m{#1}}
\newcommand{\arxivlarge}[0]{\texttt{arxiv-0923}}
\newcommand{\arxivsmall}[0]{\texttt{arxiv-0923-top40}}
\newcommand{\arxivlargep}[0]{\texttt{arxiv-0623}}
\newcommand{\arxivsmallp}[0]{\texttt{arxiv-0623-top40}}

\newcommand{\june}[0]{\texttt{arxiv-0623}}
\newcommand{\sept}[0]{\texttt{arxiv-0923}}

\newcommand{\se}[1]{\textcolor{black}{#1}}
\newcommand{\zr}[1]{\textcolor{black}{#1}}
\newcommand{\zrf}[1]{\textcolor{black}{#1}}
\newcommand{\ak}[1]{\textcolor{black}{#1}}
\newcommand{\akf}[1]{\textcolor{black}{#1}}

\title{
NLLG Quarterly arXiv Report 09/23: 
\\
What are the most influential current 
AI 
Papers?
}
\author[1]{Ran Zhang}
\author[2,1]{Aida Kostikova}
\author[1]{Christoph Leiter} 
\author[1]{Jonas Belouadi} 
\author[1]{Daniil Larionov}
\author[1]{Yanran Chen} 
\author[1]{Vivian Fresen} 
\author[1]{Steffen Eger} 
\affil[1]{Natural Language Learning Group (NLLG)\footnote{\url{https://nl2g.github.io/}}, University of Mannheim}
\affil[2]{Knowledge Representation and Machine Learning, University of Bielefeld}
\date{}

\begin{document}

\maketitle

\begin{abstract}
Artificial Intelligence (AI) has witnessed rapid growth, especially in the subfields Natural Language Processing (NLP), Machine Learning (ML) and Computer Vision (CV). Keeping pace with this rapid progress poses a considerable challenge for researchers and professionals in the field. In this arXiv report, the second of its kind, which covers the period from January to September 2023, we aim to provide insights and analysis that help navigate these dynamic 
areas of AI. We accomplish this by 1) identifying the top-40 most cited papers from arXiv in the given period, comparing the current top-40 papers to the previous report, which covered the period January to June; 2) analyzing dataset characteristics and keyword popularity; 3) examining the global sectoral distribution of institutions to reveal differences in engagement across geographical areas. Our findings highlight the continued dominance of NLP: while only 16\% of all submitted papers have NLP as primary category (more than 25\% have CV and ML as primary category), 50\% of the most cited papers have NLP as primary category, 90\% of which target LLMs. Additionally, we show that i) the US dominates among both top-40 and top-9k papers, followed by China; ii) Europe clearly lags behind and is hardly represented in the top-40 most cited papers; iii) US industry is largely overrepresented in the top-40 most influential papers.

\end{abstract}
\section{Introduction}

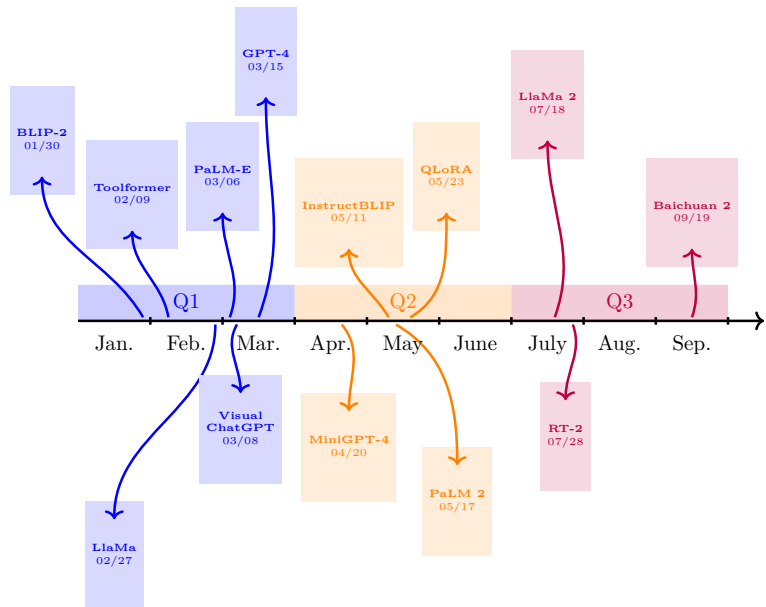
\begin{figure}[!htp]
\centering
\scalebox{.8}{\tikzstyle{descript} = [text = black,align=center, minimum height=1.8cm, align=center, outer sep=0pt,font = \tiny]
\tikzstyle{activity} =[align=center,outer sep=1pt]

\begin{tikzpicture}[very thick, black, scale=0.6]

\coordinate (O) at (-1,0); 
\coordinate (P1) at (5,0);
\coordinate (P2) at (11,0);
\coordinate (P3) at (17,0);
\coordinate (F) at (18,0); 
\coordinate (E1) at (0,0); 
\coordinate (E2) at (0.5,0); 

\fill[color=ColorOne!20] rectangle (O) -- (P1) -- ($(P1)+(0,1)$) -- ($(O)+(0,1)$); 
\fill[color=ColorTwo!20] rectangle (P1) -- (P2) -- ($(P2)+(0,1)$) -- ($(P1)+(0,1)$); 
\fill[color=ColorThree!20] rectangle (P2) -- (P3) -- ($(P3)+(0,1)$) -- ($(P2)+(0,1)$);

\draw ($(P1)+(-3,0.5)$) node[activity,ColorOne] {Q1};
\draw ($(P2)+(-3,0.5)$) node[activity,ColorTwo] {Q2};
\draw ($(P3)+(-3,0.5)$)  node[activity, ColorThree] {Q3};

\node[descript,fill=ColorOne!15, text=ColorOne](D2) at ($(O)+(-1,5)$) {%
\textbf{BLIP-2}\\ 
01/30};
\path[->,color=ColorOne] ($(O)+(1.8, 0.1)$) edge [out=120, in=-90]  ($(D2)+(0,-1)$);

\node[descript,fill=ColorOne!15,text=ColorOne](D3) at ($(P1)+(-4.5,3.5)$) {%
\textbf{Toolformer}\\
02/09};
\path[->,color=ColorOne]($(P1)+(-3.5,0.1)$)  edge [out=100, in=-90]  ($(D3)+(0,-1)$);

\node[descript,fill=ColorOne!15,text=ColorOne](D4) at ($(P1)+(-5,-6.5)$) {%
\textbf{LlaMa}\\
02/27};
\path[->,color=ColorOne]($(P1)+(-2.2,-0.1)$)  edge [out=-90, in=90]  ($(D4)+(0,1)$);

\node[descript,fill=ColorOne!15,text=ColorOne](D5) at ($(P1)+(-2, 4.)$) {%
\textbf{PaLM-E}\\
03/06};
\path[->,color=ColorOne]($(P1)+(-1.8, 0.1)$)  edge [out=70, in=-90]  ($(D5)+(0,-1)$);

\node[descript,fill=ColorOne!15,text=ColorOne](D6) at ($(P1)+(-0.8, 7.2)$) {%
\textbf{GPT-4}\\
03/15};
\path[->,color=ColorOne]($(P1)+(-1, 0.1)$)  edge [out=70, in=-90]  ($(D6)+(0,-1)$);

\node[descript,fill=ColorOne!15,text=ColorOne](D7) at ($(P1)+(-1.5, -3)$) {%
\textbf{Visual}\\
\textbf{ChatGPT}\\
03/08};
\path[->,color=ColorOne]($(P1)+(-1.6, -0.1)$)  edge [out=-120, in=90]  ($(D7)+(0, 1)$);

\node[descript,fill=ColorTwo!15,text=ColorTwo](D8) at ($(P1)+(1.5, -3.5)$) {%
\textbf{MiniGPT-4}\\
04/20};
\path[->,color=ColorTwo]($(P1)+(1.3, -0.1)$)  edge [out=-50, in=90]  ($(D8)+(0, 1)$);

\node[descript,fill=ColorTwo!15,text=ColorTwo](D9) at ($(P1)+(1.5, 3)$) {%
\textbf{InstructBLIP}\\
05/11};
\path[->,color=ColorTwo]($(P1)+(2.6, 0.1)$)  edge [out=120, in=-90]  ($(D9)+(0, -1)$);

\node[descript,fill=ColorTwo!15,text=ColorTwo](D10) at ($(P1)+(4.5, -5)$) {%
\textbf{PaLM 2}\\
05/17};
\path[->,color=ColorTwo]($(P1)+(2.8, -0.1)$)  edge [out=-30, in=90]  ($(D10)+(0, 1)$);

\node[descript,fill=ColorTwo!15,text=ColorTwo](D11) at ($(P1)+(4.2, 4)$) {%
\textbf{QLoRA}\\
05/23};
\path[->,color=ColorTwo]($(P1)+(3.2, 0.1)$)  edge [out=30, in=-90]  ($(D11)+(0, -1)$);

\node[descript,fill=ColorThree!15,text=ColorThree](D11) at ($(P2)+(1, 6)$) {%
\textbf{LlaMa 2}\\
07/18};
\path[->,color=ColorThree]($(P2)+(1.2, 0.1)$)  edge [out=70, in=-90]  ($(D11)+(0, -1)$);

\node[descript,fill=ColorThree!15,text=ColorThree](D12) at ($(P2)+(1.5, -3.2)$) {%
\textbf{RT-2}\\
07/28};
\path[->,color=ColorThree]($(P2)+(1.7, -0.1)$)  edge [out=-70, in=90]  ($(D12)+(0, 1)$);

\node[descript,fill=ColorThree!15,text=ColorThree](D11) at ($(P2)+(5, 3)$) {%
\textbf{Baichuan 2}\\
09/19};
\path[->,color=ColorThree]($(P2)+(5, 0.1)$)  edge [out=70, in=-90]  ($(D11)+(0, -1)$);



\draw[->] (O) -- (F);
\foreach \x in {1,3,...,17}
\draw(\x cm,3pt) -- (\x cm,-3pt);
\foreach \i \j in {0/Jan.,2/Feb.,4/Mar.,6/Apr.,8/May,10/June,12/July, 14/Aug., 16/Sep.}{
\draw (\i,0) node[below=3pt] {\j} ;
}

\end{tikzpicture}}
\caption{Timeline of influential models and techniques from \se{our} top-40 list based on normalized citation counts. The height of each node is related to the corresponding rank.} \label{fig:timeline}
\end{figure}

The landscape of 
Artificial Intelligence (AI) is rapidly changing, with new research emerging constantly. Th\se{is} rapid development is exemplified vividly by the timeline plot of influential models and techniques from \se{our} list of the 40 most popular papers based on normalized citation counts, illustrated in Figure \ref{fig:timeline}. Keeping up with the latest developments and understanding their impact on the field is crucial for academics \se{inside and outside of computer science}, professionals in the industry \se{as well as the wider public}. This updated report from the Natural Language Learning Group (NLLG, \url{https://nl2g.github.io/}) 
aims to serve as a guide in this dynamic environment, offering a comprehensive overview 
\se{over the (relatively) most cited}
papers on arXiv (\url{https://arxiv.org/}) from January 2023 until September 2023. \se{Our updated report puts an emphasis on additional factors besides paper content, i.e., the geographic distribution of the most influential papers.}

We continue to utilize arXiv as our primary source due to its status as a well-recognized repository for pre-print papers, as well as its quick publication process, which allows for timely access to new research. Our analysis expands upon previous efforts by incorporating an updated methodology and a broader scope that includes the arXiv subcategories of Artificial Intelligence (cs.AI) and Computer Vision and Pattern Recognition (cs.CV), in addition to our core focus on Natural Language Processing (cs.CL) and Machine Learning (cs.LG). 

The main insights from this report are: (i) NLP research remains centered on LLMs, with a slight shift toward\se{s} exploring a \se{wider} range of models beyond ChatGPT (e.g.\ LLaMA and multimodal language models); (ii) \se{we observe a} notable emphasis on LLMs in practical applications, evaluation of LLMs and multimodality involving LLMs and vision; (iii) \se{our} top-40 publications, \se{ranked by normalized citation count}, are predominantly driven by a few institutions (primarily by US companies) and we observe an equal distribution of independent and collaborative research;\footnote{Here we define independent research at affiliation level. It refers to papers whose authors \se{have} the same affiliation.} (iv) the US dominates in both industry and academia, more pronounced in industry, followed by China, \se{concerning the most influential AI papers} for both top-40 and top-9k publications while \zrf{Europe as a whole 
lacks 
presence, especially in top-40 publications;}
(v) academic research is less concentrated in specific regions for top-40 and top-9k publications \zrf{while the top-40 papers are strongly skewed towards industry compared to top-9k publications.}



\se{We} first outline 
\se{our} 
methodology in Section \ref{sec:methodology}, followed by an analysis of the dataset in Section \ref{dataset_stat}, examining the arXiv categories \se{involved} and citation patterns. Section \ref{sec:toppapers} presents the latest top-40 papers and their comparison to the top-40 list from June \cite{eger2023nllg} 
from ranking and topic perspectives. We then analyze the popularity trend of keywords and distribution of institutions by sectors across global regions. 
Section \ref{sec:conclusion} concludes our report.

Our code and data are available from \url{https://github.com/NL2G/Quaterly-Arxiv}\se{.}

\section{Methodology}\label{sec:methodology}
 We follow the methodology of our previous arXiv report \cite{eger2023nllg} to identify the most influential papers and 
 developing trend\se{s} from the AI subfields NLP, ML, and CV.
 We briefly report the methodology and highlight the differences from the previous report in the following: 
\begin{table}[!htbp]
    \centering
    \fontsize{9pt}{9pt}\selectfont
    \begin{tabular}{M{2.5cm}|M{1.5cm}M{2.5cm}M{2cm}M{4cm}}
         \hline
         \textbf{Dataset name} & \textbf{Size} & \textbf{Time period} & \textbf{\#Primary Categories} & \textbf{Search Query}\\ 
         \midrule
         \arxivlarge &  47\se{,}331 & \multirow{2}{*}{01/01-09/30 (2023)} & 130 & \multirow{2}{*}{cs.CL, cs.LG, cs.AI, cs.CV}\\
         \arxivsmall &  40 & & 6 &\\
         \hdashline
         \arxivlargep &  20,843 & \multirow{2}{*}{01/01-06/31 (2023)} & 123 & \multirow{2}{*}{cs.CL, cs.LG}\\
         \arxivsmallp &  40 & & 5\\
         \bottomrule
    \end{tabular}
    \caption{Statistics on our released datasets. Size is the number of papers in each dataset; \#Primary Categories gives the number of distinct primary arXiv categories our papers are assigned to. A 
    paper is retrieved if it includes any of the categories listed in Search Query.}
    \label{table:stats}
\end{table}
 
\begin{enumerate}
\item \textbf{Data Retrieval from arXiv}: 
    We collect all papers from 01/01/2023 to 09/30/2023 belonging to the arXiv categories cs.CL (computation and language), cs.AI (artificial intelligence), cs.CV (computer vision and pattern recognition) and cs.LG (machine learning) using a Python arXiv API.\footnote{ArXiv papers may belong to several categories. We retrieve 
    \se{a} 
    paper if it includes any of the four categories \se{listed}.} 
    Our retrieval time is \textbf{October 26, 2023}. The retrieval time is important because citation counts are constantly in flux. Since arXiv papers can be updated anytime, we take the first submission date of a paper to arXiv as its publication date.

\item \textbf{Z-score calculation:}
 For each paper, we extract its citation count from Semantic Scholar \url{https://www.semanticscholar.org/} as a measure of popularity and arguably importance \cite{aksnes2019citations}. To reduce the time effect on citation counts (e.g.\ papers that are published earlier have a higher chance to be cited than recent papers), we calculate a \emph{normalized citation count} by determining \emph{how many standard deviations a paper is above the mean of citations of all papers published in the same week}. This is the so-called z-score of Newman \cite{newman2014prediction}:
 \begin{align}
    z_{\se{i}(t)} = \frac{c_{\se{i}(t)}-\textit{mean}(\mathbf{c}(t))}{\textit{std}(\mathbf{c}(t))} \label{eq:eq1}
 \end{align}
 for a paper $i$ published in week $t$ with citation count $c_{\se{i}(t)}$; $\mathbf{c}(t)$ is the list of citation counts of all papers published in week $t$. 

We interpret the z-score 
\se{as follows}: compared to all papers published in the same week, if a paper lies several standard deviations \textbf{above} the mean, it can be considered excellent compared to its peers. For example, in a normal distribution, only about 16\% of data points lie one standard deviation above the mean value. 

\textbf{Stable z-score for final ranking\se{:}}
In our previous 
\se{report,} 
we split the weeks by the first day of the year 2023 (i.e.\ \emph{Sunday-Saturday}). To remove the dependence on a specific week split and to reduce the effect of different week splits on z-score calculation for the current and next reports, we compute the averaged z-score for each paper over all possible week splits (in total 
7 splits e.g.\ Sunday-Saturday, Monday-Sunday, and so on). If the standard deviation of a paper's z-scores from all possible splits remains low, it indicates that the resulting averaged z-score is more stable throughout all time windows. Therefore, we deduct the standard deviation from the averaged z-score to punish instability:\footnote{\zrf{This value is also used in other studies e.g.\ as the threshold for value cut-off \cite{schmitt1998quantifying} or as the threshold for indicating ``worst case scenario'' in clinical studies \cite{jakobsen2014thresholds} and social science \cite{krause1986social}.}} 
 \begin{align}
     \hat{z}_{i\se{(t)}}= \textit{mean}(\mathbf{z}_{i(t^d)}) - \textit{std}(\mathbf{z}_{i(t^d)}) \label{eq:eq2}
 \end{align}
 for paper $i$ published in week $t$, where ${t^d}$ 
 \se{indexes}
 different week divisions starting with weekday 
$d \in \{\text{Mon, Tue, Wed, Thu, Fri, Sat, Sun}\}$. $\mathbf{z}_{i(t^d)}$ is a list of z-scores computed under \se{the} different week splits using Equation (\ref{eq:eq1}).

\item \textbf{Manual Evaluation\se{:}}
We manually assess papers in the top-40 list by cross-checking the publication date on arXiv with their actual first publication/release/submission date. If a paper is available earlier, we exclude it from our evaluation.

\end{enumerate}

Steps 1 and 2+3 result in two distinct datasets that we release with this report. We refer to the dataset obtained from step 1 as \arxivlarge{}. Since we add two more categories to our search query, we obtain a dataset that doubles the size of the previous dataset \arxivlargep{}. 
We refer to the dataset resulting from step 2+3, \se{after sorting the papers by z-score and capping at the $n=40$ top papers}, as \arxivsmall{}. This dataset contains the top-40 
papers ranked by stable z-score. Table \ref{table:stats} gives elementary statistics on each of the \se{two datasets} in comparison to the previous datasets.

\begin{figure}[!htb]
    \centering
        \includegraphics[width=0.5\textwidth]{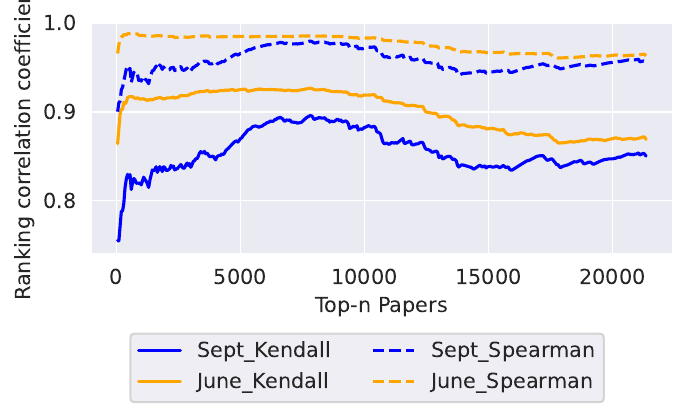}
        \caption{Distribution of Kendall’s tau and Spearman ranking correlation coefficient of scores over top-n papers computed using Equation (\ref{eq:eq1}) with a single week division (i.e.\ Sunday-Saturday, same as the previous report) vs.\ stable z-scores using Equation 
      (\ref{eq:eq2}) using all week divisions. We compute the correlation coefficient using both datasets \arxivlargep{} (citation count in \textit{June}) and \arxivlarge{} (citation count in \textit{Sept}).}
    \label{fig:rank_corr}
    \hfill
\end{figure}

\paragraph{\zrf{The effect of old vs.\ new z-score formula.}}
In order to examine the \se{effect of 
our new formula, i.e.\ Equation (\ref{eq:eq2}),} we compute the ranking correlation of  
\se{the} z-scores computed with 1) a single week division using Equation (\ref{eq:eq1}) (i.e.\ Sunday-Saturday, consistent with the previous report) and 2) stable z-score using Equation (\ref{eq:eq2}) (i.e.\ considering all week divisions). Figure \ref{fig:rank_corr} shows the distribution of the correlation coefficient between both scores over top-n papers (ranked by stable z-score). For both datasets, we observe a high \se{rank} correlation between the two scores. \se{For example}, if we compare the \june{} top-40 lists 
ranked by 
\se{both} scores, 36 out of 40 papers appear in both lists \se{(even though some of their ranks may have chaned)} with 4 new papers at rank 35-37 and 39. 

Computing the average z-score over all possible week splits (Sunday-Saturday, Monday-Sunday, etc.) is a robust way to remove dependence on how a week is defined. This approach ensures that the z-score is not overly influenced by the particular day a paper was published within the week.



\section{Dataset statistics}
\label{dataset_stat}
\paragraph{How many arXiv categories (scientific subfields) are involved?}
Our latest dataset \arxivlarge{} comprises 47,331 papers submitted to arXiv between 01/01/2023 and 09/30/2023 with at least one of the indicated categories given as cs.CL, cs.LG, cs.AI or cs.CV. Given that NLP, ML and CV have a 
\se{noticable} 
impact across various domains, it is often that these papers are not exclusively sourced from one category. \se{Indeed,} our paper collection is distributed across 130 distinct primary arXiv categories. We give detailed statistics on primary categories occurring at least 100 times in Table \ref{table:categories} in the appendix. The most frequent top-level categories are cs (computer science), stat (statistics), eess (electrical engineering and systems science), math (mathematics), quant-ph (quantum physics) and q-bio (quantitative biology). Apart from the four fine-grained categories we utilize in our data query (i.e., cs.CV, cs.LG, cs.CL, cs.AI), eess.IV (image and video processing), stat.ML (statistics, machine learning) and cs.RO (robotics) are also frequent primary categories.  
\begin{figure}[!htb]
     \centering
     \begin{subfigure}[b]{0.48\textwidth}
         \centering
         \includegraphics[width=\textwidth]{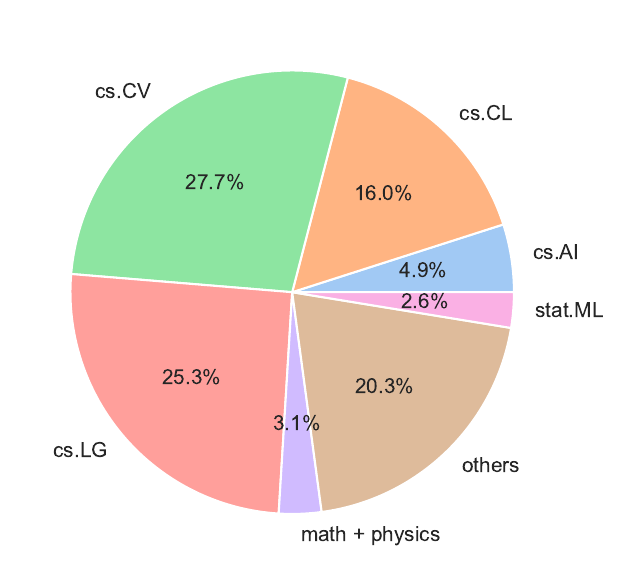}
         \caption{Pie chart of distribution of main categories in our dataset.}
        \label{fig:categories}
     \end{subfigure}
     \hfill
     \begin{subfigure}[b]{0.48\textwidth}
         \centering
         \includegraphics[width=\textwidth]{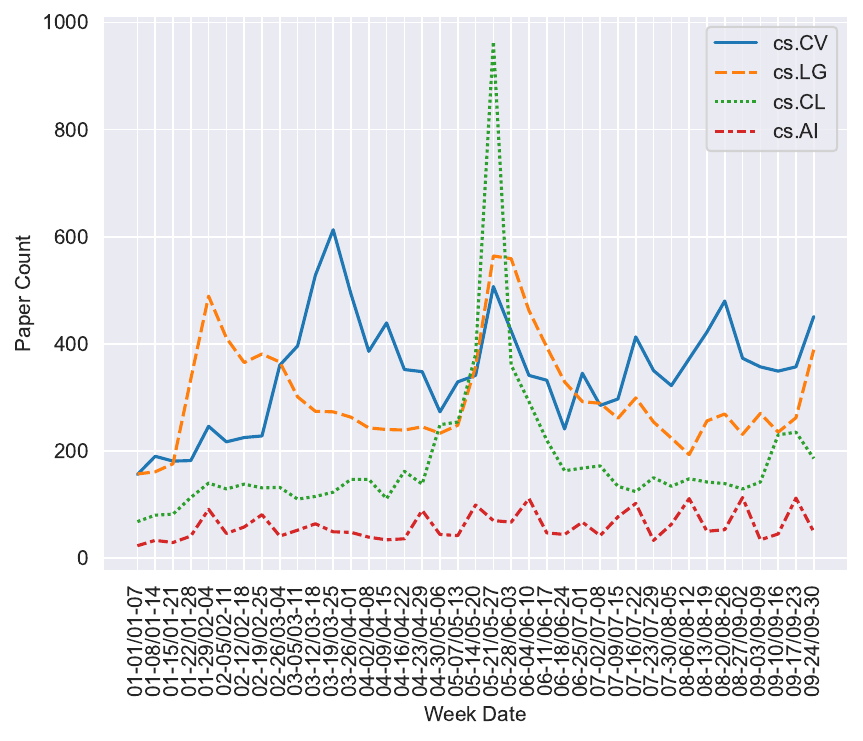}
            \caption{Distribution of weekly submission count by primary categories.}
        \label{fig:weekly}
     \end{subfigure}
     \caption{Distribution plots by main categories and by weekly submission.}
     \label{fig:dataset_plot}
\end{figure}

Figure \ref{fig:categories} is a pie chart summarizing the distribution of primary categories. We merge the top-level categories from math and physics into one main category math+physics. \se{The categories} \se{c}s.LG with 25.3\% and cs.CV with 27.7\% of papers are the two largest categories.
\se{The NLP category} \se{c}s.CL ranks third with 16.0\%. However, it dominates the top-40 list and all top 5 papers belong to this category. We also notice that 3.1\% of papers belong to the top-level category of math or physics, which indicates the interdisciplinary influence of AI-related methods in science subjects.    

Figure \ref{fig:weekly} shows paper \se{submission} counts per category over the weeks. We observe 
periodic jump\se{s} in submission counts for cs.CV and cs.LG. Mostly, the jump\se{s} occur 
in the third week of each month. For cs.CV, such turbulence occurs more often (nearly every month), while for cs.LG and cs.CL, 
\se{a} 
big jump in submission count occurs in May. To inspect the reason behind the \textbf{spike of cs.CL in May}, we analyze the comments of papers submitted in the week from 05-21 to 05-27 where 526 out of 964 papers contain author comments. As the week marks the 2023 camera-ready submission deadline for the popular NLP conference ACL (\url{https://www.aclweb.org/portal/content/acl-2023-call-papers}), we start evaluating ACL-related keywords. 
\se{In fact,} 
24.7\% of papers (238 out of 964) mention ``acl'' in their comments where these papers are accepted to the main conference, findings, or workshop of ACL 2023. Notably, we identify 44 papers mentioning EMNLP. One reason might be that 05-27 is the ARR review deadline for EMNLP submission (\url{https://2023.emnlp.org/calls/main_conference_papers/}). This may also be due to the unique submission policy for NLP conferences, i.e.\ one month anonymity period. We use the same method to study the spike of cs.CV submission in March. Evidence shows that the high number of submissions in March is linked to the camera-ready submission deadline of the popular CV conference CVPR which is 24th of March. In this week, 194 out of 613 papers (31.6\%) mention CVPR acceptance in the comments.

\paragraph{How many citations and standard deviations are there per week \se{for} each \se{c}ategory?}
\begin{figure}[!htb]
    \centering
    \includegraphics[width=0.8\textwidth]{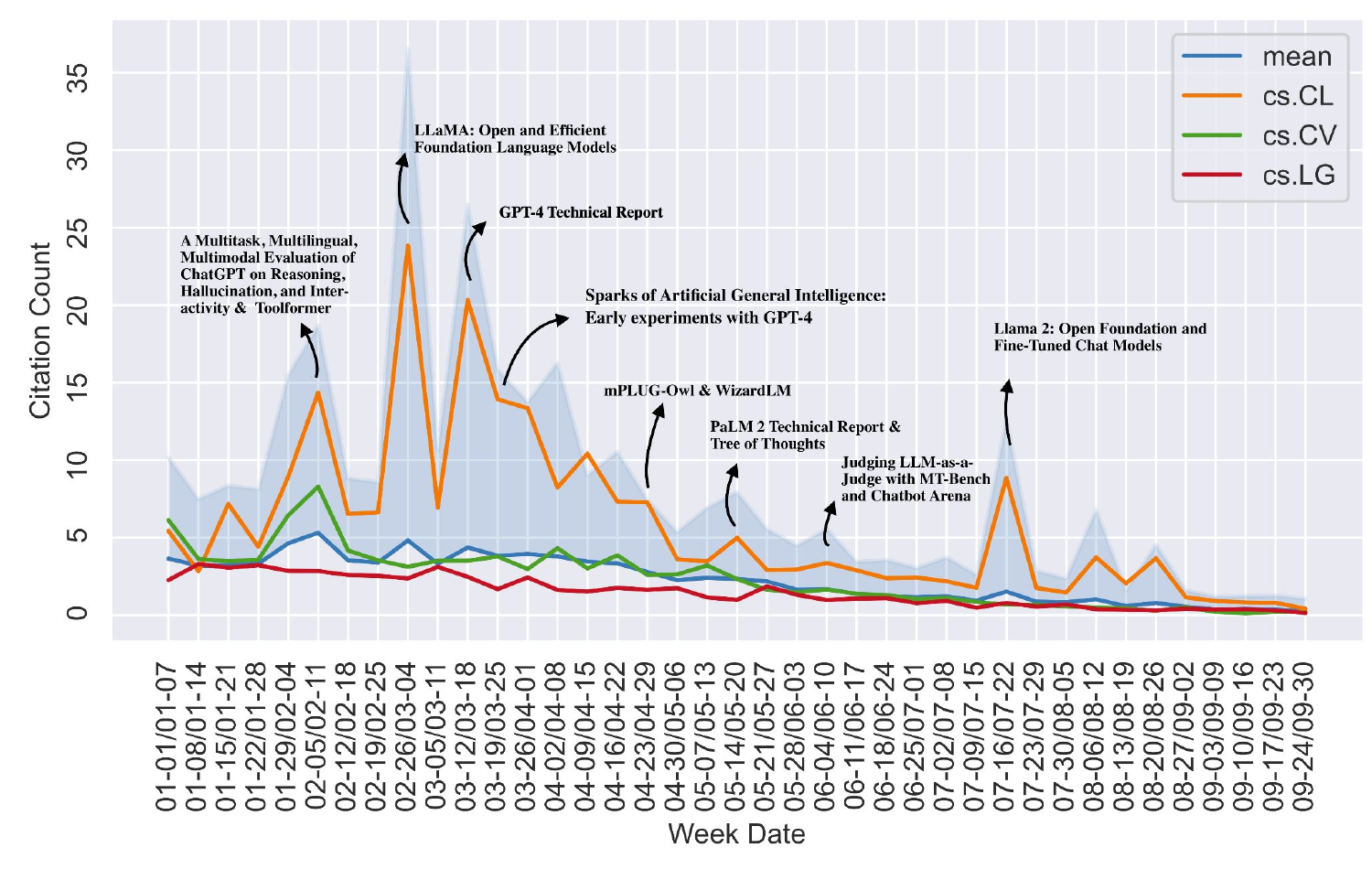}
    \caption{Distribution of citation mean over all papers\se{,} including 0.5 standard deviations as upper bound (light blue region), and for the primary categories cs.CL, cs.LG and cs.CV. We mark some influential papers from category cs.CL.}
    \label{fig:cit_count}
\end{figure}
Figure \ref{fig:cit_count} presents mean citation counts for papers in several categories over the observed period marked with influential papers. For more information, please see Table \ref{table:mean} (in the appendix). 
We draw the following insights from this:
\begin{itemize}
\item Over time, citation counts tend to decrease, with papers averaging fewer than two citations by May across all categories, reflecting a pattern in which newer research initially receives fewer citations.

\item The cs.CL category consistently surpasses cs.LG and other categories in average citations, indicating a higher interest and activity in NLP-related research, despite the fact that cs.CL lacks behind cs.LG and cs.CV in submission counts, see Figures \ref{fig:cit_count} and \ref{fig:weekly}.

\item In February, particularly within cs.CL, notable spikes in citation numbers are observed. These include \se{highly cited papers} 
such as \textit{A Multitask, Multilingual, Multimodal Evaluation of ChatGPT on Reasoning, Hallucination, and Interactivity} \cite{bang2023multitask} and Toolformer \cite{schick2023toolformer} in the second week, {followed by the release of} the LLaMA model \cite{touvron2023llama} in the last week. 

Moreover, a surge in the \zrf{mean} \se{citation numbers} 
during the week of March 12th aligns with the publication of the \textit{GPT-4 Technical Report} \cite{openai2023gpt4}, as well as \textit{Sparks of Artificial General Intelligence: Early experiments with GPT-4} \cite{bubeck2023sparks}, which discusses its rising capabilities and implications.

\item Mid-July also sees a notable increase in citations, coinciding with the submission of LLama 2 \cite{touvron2023llama2}.

\item Although citations in cs.CV are  
\se{fewer} than in cs.CL, this category also demonstrates some surges in citations, aligning with the publication of key papers, such as \textit{BLIP-2} (week of 29 January), followed by \textit{Adding Conditional Control to Text-to-Image Diffusion Models} \cite{zhang2023adding} in the week of February 5.
\end{itemize}
 
Interestingly, despite the downward trend in the number of citations towards the end of the period, there are instances of increased variability in citation counts, such as in the week of August 6, where the standard deviation notably rises to 11.55 against a mean of just 1.03 (see Table \ref{table:mean}). 
In this case, the high standard deviation relative to the mean indicates that although most papers have received citations close to the average, some papers have much higher citation counts. The corresponding z-score would reflect such characteristics.


 \begin{table}[!htbp]
 \centering
\fontsize{9.5pt}{9.5pt}\selectfont
 \begin{tabular}{|M{0.5cm}|m{5.5cm}|M{0.9cm}|M{3cm}|M{1.8cm}|M{0.5cm}|M{0.7cm}|M{0.4cm}|}
\hline
\textbf{No.}& \multicolumn{1}{c|}{\textbf{Title}} & \textbf{Cat.}& \textbf{Link} & \textbf{Week} &\textbf{Cit} & \textbf{z-score} & \textbf{\text{$\uparrow \downarrow$}} \\ 
\hline
1 &GPT-4 Technical Report &cs.CL&\url{http://arxiv.org/abs/2303.08774v3} &03-12/03-18 &1573 &35.1 &\textcolor{green}{\text{$\uparrow$}1} \\
2 &\textbf{Llama 2: Open Foundation and Fine-Tuned Chat Models} &cs.CL&\url{http://arxiv.org/abs/2307.09288v2} &07-16/07-22 &778 &34.3 & \\
3 &LLaMA: Open and Efficient Foundation Language Models &cs.CL&\url{http://arxiv.org/abs/2302.13971v1} &02-26/03-04 &2243 &33.6 &\textcolor{red}{\text{$\downarrow$}2} \\
4 &Sparks of Artificial General Intelligence: Early experiments with GPT-4 &cs.CL&\url{http://arxiv.org/abs/2303.12712v5} &03-19/03-25 &819 &32.5 &\textcolor{red}{\text{$\downarrow$}1} \\
5 &Judging LLM-as-a-Judge with MT-Bench and Chatbot Arena &cs.CL&\url{http://arxiv.org/abs/2306.05685v3} &06-04/06-10 &262 &32.3 &\textcolor{green}{\text{$\uparrow$}6} \\
6 &\textit{BLIP-2: Bootstrapping Language-Image Pre-training with Frozen Image Encoders and Large Language Models }&cs.CV&\url{http://arxiv.org/abs/2301.12597v3} &01-29/02-04 &635 &27.9 & \\
7 &Segment Anything &cs.CV&\url{http://arxiv.org/abs/2304.02643v1} &04-02/04-08 &721 &26.0 &\textcolor{green}{\text{$\uparrow$}3} \\
8 &PaLM 2 Technical Report &cs.CL&\url{http://arxiv.org/abs/2305.10403v3} &05-14/05-20 &299 &24.6 &\textcolor{red}{\text{$\downarrow$}4} \\
9 &The RefinedWeb Dataset for Falcon LLM: Outperforming Curated Corpora with Web Data, and Web Data Only &cs.CL&\url{http://arxiv.org/abs/2306.01116v1} &05-28/06-03 &135 &23.1 &\textcolor{green}{\text{$\uparrow$}19} \\
10 &PaLM-E: An Embodied Multimodal Language Model &cs.LG &\url{http://arxiv.org/abs/2303.03378v1} &03-05/03-11 &345 &22.4 &\textcolor{red}{\text{$\downarrow$}5} \\
11 &QLoRA: Efficient Finetuning of Quantized LLMs &cs.LG &\url{http://arxiv.org/abs/2305.14314v1} &05-21/05-27 &186 &21.8 &\textcolor{red}{\text{$\downarrow$}2} \\
12 &Visual Instruction Tuning &cs.CV&\url{http://arxiv.org/abs/2304.08485v1} &04-16/04-22 &300 &19.8 &\textcolor{red}{\text{$\downarrow$}5} \\
13 &\textbf{Universal and Transferable Adversarial Attacks on Aligned Language Models }&cs.CL&\url{http://arxiv.org/abs/2307.15043v1} &07-23/07-29 &86 &19.6 & \\
14 &Tree of Thoughts: Deliberate Problem Solving with Large Language Models &cs.CL&\url{http://arxiv.org/abs/2305.10601v1} &05-14/05-20 &237 &19.5 &\textcolor{red}{\text{$\downarrow$}1} \\
15 &\textit{Adding Conditional Control to Text-to-Image Diffusion Models }&cs.CV&\url{http://arxiv.org/abs/2302.05543v2} &02-05/02-11 &546 &19.2 & \\
16 &A Survey of Large Language Models &cs.CL&\url{http://arxiv.org/abs/2303.18223v12} &03-26/04-01 &482 &18.5 &\textcolor{red}{\text{$\downarrow$}8} \\
17 &\textit{MiniGPT-4: Enhancing Vision-Language Understanding with Advanced Large Language Models} &cs.CV&\url{http://arxiv.org/abs/2304.10592v2} &04-16/04-22 &277 &18.1 & \\
18 &\textbf{Large Language Models as Optimizers} &cs.LG &\url{http://arxiv.org/abs/2309.03409v1} &09-03/09-09 &31 &17.8 & \\
19 &Voyager: An Open-Ended Embodied Agent with Large Language Models &cs.AI&\url{http://arxiv.org/abs/2305.16291v2} &05-21/05-27 &124 &17.3 &\textcolor{red}{\text{$\downarrow$}7} \\
20 &\textbf{Lost in the Middle: How Language Models Use Long Contexts }&cs.CL&\url{http://arxiv.org/abs/2307.03172v2} &07-02/07-08 &90 &17.1 & \\
\hline
 \end{tabular}
 \caption{Papers, their prime category, arXiv link, week of first arXiv submission, citation count (as of 10/27/2023) and \textbf{stable z-score}. \textbf{Top 20 papers} according to \textbf{stable z-score} among all 
 \arxivlarge{} papers. Papers published from June to September are shown in bold text. Papers published before June but appear first time in the list are shown in italic text.}
 \label{table:top20}
 \end{table}
\begin{table}[!htbp]
 \centering
 \fontsize{9.5pt}{9.5pt}\selectfont
 \begin{tabular}{|M{0.5cm}|m{5.5cm}|M{0.9cm}|M{3cm}|M{1.8cm}|M{0.5cm}|M{0.7cm}|M{0.4cm}|}
\hline
\textbf{No.}& \multicolumn{1}{c|}{\textbf{Title}} & \textbf{Cat.}& \textbf{Link} & \textbf{Week} &\textbf{Cit} & \textbf{z-score} & \textbf{\text{$\uparrow \downarrow$}} \\ 
 \hline
21 & InstructBLIP: Towards General-purpose Vision-Language Models with Instruction Tuning &cs.CV&\url{http://arxiv.org/abs/2305.06500v2} &05-07/05-13 &178 &16.5 &\textcolor{green}{\text{$\uparrow$}12} \\
22 &\textit{Direct Preference Optimization: Your Language Model is Secretly a Reward Model} &cs.LG &\url{http://arxiv.org/abs/2305.18290v1} &05-28/06-03 &96 &14.8 & \\
23 &\textbf{The Rise and Potential of Large Language Model Based Agents: A Survey }&cs.AI&\url{http://arxiv.org/abs/2309.07864v3} &09-10/09-16 &25 &14.7 & \\
24 &\textbf{RT-2: Vision-Language-Action Models Transfer Web Knowledge to Robotic Control }&cs.RO &\url{http://arxiv.org/abs/2307.15818v1} &07-23/07-29 &63 &14.5 & \\
25 &\textbf{Siren's Song in the AI Ocean: A Survey on Hallucination in Large Language Models} &cs.CL&\url{http://arxiv.org/abs/2309.01219v2} &09-03/09-09 &28 &14.5 & \\
26 &A Multitask, Multilingual, Multimodal Evaluation of ChatGPT on Reasoning, Hallucination, and Interactivity &cs.CL&\url{http://arxiv.org/abs/2302.04023v2} &02-05/02-11 &406 &13.8 &\textcolor{red}{\text{$\downarrow$}20} \\
27 &StarCoder: may the source be with you! &cs.CL&\url{http://arxiv.org/abs/2305.06161v1} &05-07/05-13 &130 &13.7 &\textcolor{green}{\text{$\uparrow$}9} \\
28 &mPLUG-Owl: Modularization Empowers Large Language Models with Multimodality &cs.CL&\url{http://arxiv.org/abs/2304.14178v1} &04-23/04-29 &131 &13.6 &\textcolor{green}{\text{$\uparrow$}4} \\
29 &\textbf{A Survey on Evaluation of Large Language Models }&cs.CL&\url{http://arxiv.org/abs/2307.03109v8} &07-02/07-08 &72 &13.6 & \\
30 &Large Language Models are not Fair Evaluators &cs.CL&\url{http://arxiv.org/abs/2305.17926v2} &05-28/06-03 &88 &13.5 &\textcolor{red}{\text{$\downarrow$}1} \\
31 &ImageBind: One Embedding Space To Bind Them All &cs.CV&\url{http://arxiv.org/abs/2305.05665v2} &05-07/05-13 &124 &13.1 &\textcolor{red}{\text{$\downarrow$}14} \\
32 & Toolformer: Language Models Can Teach Themselves to Use Tools &	cs.CL&	\url{http://arxiv.org/abs/2302.04761v1}	& 02-05/02-11&	366&	13.0&	\textcolor{red}{\text{$\downarrow$}16}\\
33 &\textit{Visual ChatGPT: Talking, Drawing and Editing with Visual Foundation Models} &cs.CV&\url{http://arxiv.org/abs/2303.04671v1} &03-05/03-11 &207 &12.7 & \\
34 &\textit{WizardLM: Empowering Large Language Models to Follow Complex Instructions} &cs.CL&\url{http://arxiv.org/abs/2304.12244v2} &04-23/04-29 &147 &12.5 & \\
35 &\textbf{Large Language Models} &cs.CL&\url{http://arxiv.org/abs/2307.05782v2} &07-09/07-15 &57 &12.3 & \\
36 &\textit{DINOv2: Learning Robust Visual Features without Supervision }&cs.CV&\url{http://arxiv.org/abs/2304.07193v1} &04-09/04-15 &181 &12.0 & \\
37 &\textbf{Baichuan 2: Open Large-scale Language Models} &cs.CL&\url{http://arxiv.org/abs/2309.10305v2} &09-17/09-23 &24 &11.9 & \\
38 & A \textit{Comprehensive Survey on Pretrained Foundation Models: A History from BERT to ChatGPT} &cs.AI&\url{http://arxiv.org/abs/2302.09419v3} &02-12/02-18	&148&	11.9 &\\
39	& \textit{Let's Verify Step by Step}	& cs.LG	& \url{http://arxiv.org/abs/2305.20050v1} & 05-28/06-03 & 69 &	11.4	& \\
40 & \textit{Otter: A Multi-Modal Model with In-Context Instruction Tuning} & cs.CV &\url{http://arxiv.org/abs/2305.03726v1} &04-30/05-06 &91 &10.8 &\\
\hline
 \end{tabular}
 \caption{Papers, their prime category, arXiv link, week of first arXiv submission, citation count (as of 07/29/2023) and \textbf{stable z-score}. \textbf{Papers 21-40} according to \textbf{stable z-score} among all 
 \arxivlarge{} papers. Papers published from June to September are shown in bold text. Papers published before June but appear first time in the list are shown in italic text.}
 \label{table:top40}
 \end{table}

\section{Top-N papers}\label{sec:toppapers}
Table\se{s} \ref{table:top20} and \ref{table:top40} showcase the top-40 papers extracted according to the methodology described in Section \ref{sec:methodology}. 
Figure \ref{fig:old-new} shows the 
distribution \se{of categories} 
\se{in the} 
top-40 lists 
\se{for} \june{} and \sept{}.
Overall, we find that:
\begin{itemize}
\begin{figure}
    \centering
          \includegraphics[width=0.6\textwidth]{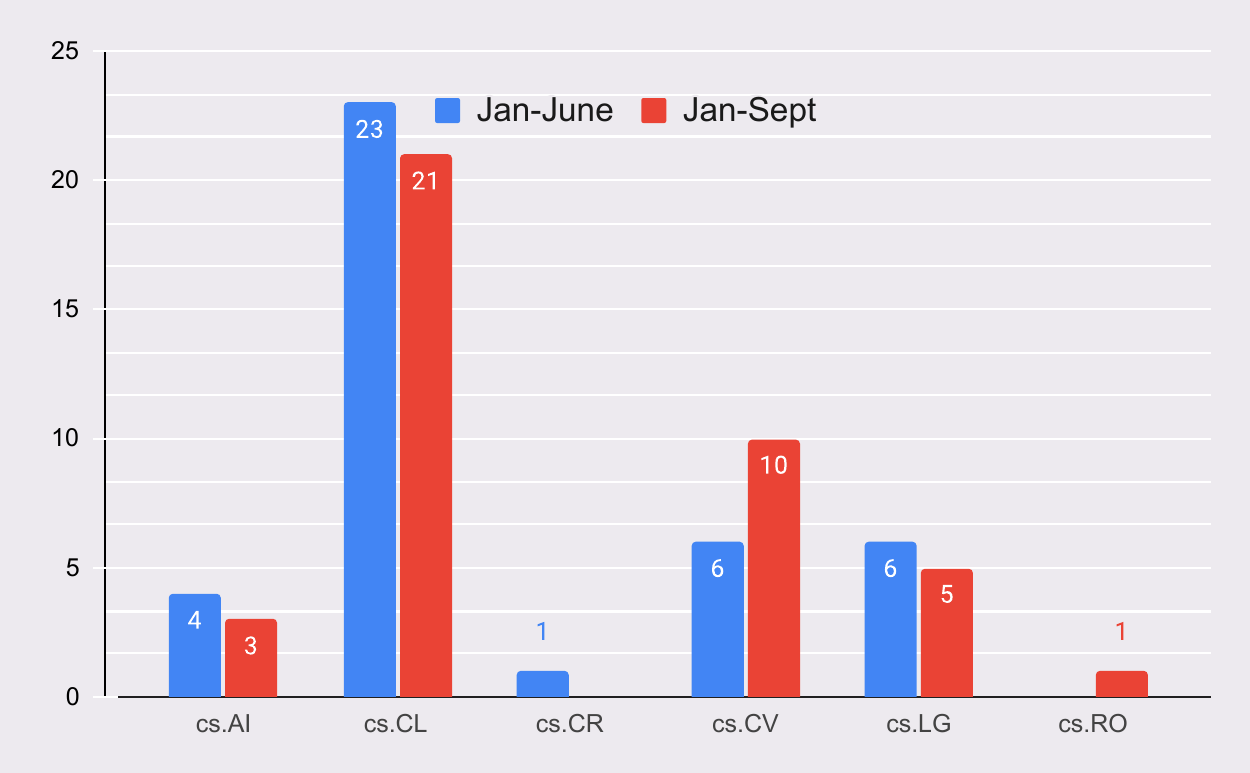}
    \caption{Distribution of papers by category from \sept{} top-40 list (Tables \ref{table:top20} and \ref{table:top40}) \& \june{} top-40 list (Tables \ref{table:top20june} and \ref{table:top40june}). }
    \label{fig:old-new}
\end{figure}
\item \textbf{\se{The} primary arXiv category cs.CL \textit{again} dominates the list} with 21 out of 40 \se{papers} in total (11 in the top-20 list, see Table \ref{table:top20}; 10 in the 20-40 list, see Table \ref{table:top40}). The second \se{largest} contributor is cs.CV \se{with 10 papers}, 5 \se{in} the top-20 list and 5 \se{in} the 20-40 list. This number increases by 4 papers compared to the top-40 list from \june{}. This may be \se{because} we add cs.CV into our search query during the latest data collection procedure. 5 papers from cs.LG are in the current top 40-list. Among them, 3 papers make it to the top 20. 
The rest \se{of the} papers in the top-40 list come from cs.AI (3 times) and cs.RO (robotics; 1 time).  

\item \textbf{The absolute citation counts vary drastically in our top-40 list. }The lowest count from the top-40 list is 24 from \se{a} 
paper published in late September (Baichuan 2: Open Large-scale Language Models \cite{yang2023baichuan}) whereas in our top-20 list, 31 is the lowest citation count for \emph{Large Language Models as Optimizers} \cite{yang2023large} published in early September. In contrast, the highest count \se{amounts} to 2243 for the paper \emph{LLaMA: Open and Efficient Foundation Language Models} \cite{touvron2023llama} published in mid-March followed by \emph{GPT-4 Technical Report} \cite{openai2023gpt4} \se{with 1573 citations}.   

\end{itemize}

Moving forward with the detailed analysis of the top-N list papers, we first discuss the changes in the top-N list. Subsequently, we outline topic-specific changes. 

\subsection{
\se{Changes from the \june{} report}
}

\begin{table}[H]
\fontsize{9pt}{9pt}\selectfont
\centering
\begin{tabular}{c cccc c cccc c}\toprule
\textbf{Publication time}&\multicolumn{4}{c}{\textbf{Jan-June}} & & \multicolumn{4}{c}{\textbf{July-Sept}} & \multirow{2}{*}{\centering\textbf{Sum}}\\
\cmidrule{1-5}\cmidrule{6-10}
\textbf{Primary Category}&cs.CL &cs.CV &cs.LG &cs.AI & &cs.CL &cs.LG &cs.AI &cs.RO & \textbf{Total}\\\cmidrule{2-5}\cmidrule{7-10}
\textbf{top-20} & &3 (3) & & & &3 &1 & & &7 \\
\textbf{21-40} &1 &3 (2)&2 &1 & &4 & &1 &1 &13 \\
\bottomrule
\end{tabular}
\caption{\label{table:old-new}
New papers that appear in the \sept{} top-40 list (Tables \ref{table:top20} and \ref{table:top40}) compared to the \june{} top-40 list (Tables \ref{table:top20june} and \ref{table:top40june} in the appendix). \zrf{We list the number of new papers in the top-20 and the 21-40 list by their primary categories and the publication time during the period of \textit{Jan-June} and \textit{July-Sept}. In the bracket, we list the number of papers that are newly introduced to the list solely because of changes in search query i.e.\ papers containing subcategories cs.CV or cs.AI without cs.CL and cs.LG. }}
\end{table}
\paragraph{Which papers are new in the top-N list?}
Table \ref{table:old-new} shows the count of new papers that appear in the \sept{} top-40 list (Tables \ref{table:top20} and \ref{table:top40}) compared to the \june{} top-40 list (Tables \ref{table:top20june} and \ref{table:top40june} in the appendix). Overall, we observe \textbf{20 new papers} that were not available in the \june{} top-40 list\se{:} 
\se{7 in the top-20 list and 13 in the 21-40 list}. 
Among the 7 papers that newly appear in the top-20 list, \zrf{3 papers with the primary category cs.CV are published before July and do not belong to \june{} dataset (without cs.CL or cs.LG in their subcategories). These papers are included in the list solely due to the change of the search query.} They all focus on \textbf{multimodal} research including \textit{BLIP-2} \cite{li2023blip}, \textit{text-to-image diffusion model\se{s}} \cite{zhang2023adding} and \textit{MiniGPT-4} \cite{zhu2023minigpt}.  
The four new papers published after June include a \textbf{technical report} for Llama 2 \cite{touvron2023llama2}, two papers addressing LLMs' vulnerabilities and limitations (i.e.\ \textit{Universal and transferable adversarial attacks on aligned language models} \cite{zou2023universal} and \textit{Lost in the Middle: How Language Models Use Long Contexts} \cite{liu2023lost}) and one paper leveraging potentials of LLMs targeting \textbf{problem-solving} (e.g.\ \textit{Large Language Models as Optimizers} \cite{yang2023large}).

The 21-40 list undergoes more changes. There are 13 new papers that were not present in the \june{} top-40 list. Among them, 7 are published before July. \zrf{2 papers belonging to cs.CV without cs.LG or cs.CL in subcategories, i.e.\ \textit{Visual ChatGPT} \cite{wu2023visual} and \textit{DINOv2} \cite{oquab2023dinov2}, were not available in the \june{} dataset. Other papers from \june{} entering the top-40 list include} 
papers concerning \se{(1)} multimodality (\textit{Otter: A Multi-Modal Model with In-Context Instruction Tuning} \cite{li2023otter}); (2) the application of LLMs focusing on instruction data generation \textit{WizardLM} \cite{xu2023wizardlm} and LLMs' output controlling \textit{Direct preference optimization} \cite{rafailov2023direct}; (3) LLMs for reasoning \textit{Let's Verify Step by Step} \cite{lightman2023let}; (4) survey on foundation models \cite{zhou2023comprehensive}. 
Meanwhile, the 21-40 list contains 6 new papers published after June. Four new \textbf{survey papers} enter the list discussing hallucination of LLMs \zrf{\textit{Siren’s Song in the AI Ocean}}\cite{zhang2023siren}, 
evaluation of LLMs \cite{chang2023survey}, LLM based agents \cite{xi2023rise}, and foundational concepts behind LLMs (\textit{Large Language Models} \cite{douglas2023large} which contains lectures for readers with a background in mathematics or physics). The topic of \textbf{multimodality} is expanded by a new paper: \textit{RT-2 model} \cite{brohan2023rt}, which integrates the vision, language, and action of robots. There is also one open source multilingual LLMs \textbf{technical report}, \textit{Baichuhan 2} \cite{yang2023baichuan}.

\paragraph{How does the rank of the papers change?}
The last column\se{s} in Table\se{s} \ref{table:top20} and \ref{table:top40} indicate 
ranking changes of top papers compared to 
\se{the 
\june{} top-40 list in Tables \ref{table:top20june} and \ref{table:top40june}}. 
\textit{The RefinedWeb Dataset for Falcon LLM} 
\cite{penedo2023refinedweb} has moved up 19 spots to \se{the} 9th \se{place}, marking the most improved paper from \june{} list. Another paper\se{,} \textit{StarCoder} \cite{li2023starcoder}\se{,} 
also moves up 9 spots. Both works highlight the power of leveraging \textbf{large-scale} datasets for training models, with \textit{The RefinedWeb Dataset} challenging the necessity of curated data for LMs and \textit{StarCoder} utilizing a vast collection of GitHub code for code LLMs. In terms of ranking decline, the paper \textit{A Multitask, Multilingual, Multi-modal Evaluation of ChatGPT on Reasoning, Hallucination, and Interactivity} \cite{bang2023multitask} marks the largest decrease in ranking by 20 spots to 
\se{rank} 26 in the new list. ImageBind \cite{girdhar2023imagebind} and Toolformer \cite{schick2023toolformer} also experience a decline by 14 and 16 spots\se{,} respectively. This coincides with a shift of interest away from ChatGPT as well as the entry of more recent papers into the top-40 list. 

\subsection{\se{What are the top-40 papers about?}
}
\begin{table}[H]
\fontsize{9pt}{9pt}\selectfont
\centering
\begin{tabular}{ccc}
\toprule
& \multicolumn{2}{c}{Topic Distribution} \\
\cmidrule{2-3}
Topic & \june{} list (Top 20 / 21-40) & \sept{} list (Top 20 / 21-40) \\
\midrule
Technical report & 4 / 3 & 6 / 5 \\
LLMs & 17 / 17 & 18 / 18 \\
Multimodality & 3 / 5 & 5 / 6 \\
Problem-solving with LLM & 2 / 0 & 2 / 4 \\
Benchmarking, Evaluation & 1 / 4 & 3 / 1 \\
Survey paper & 1 / 2 & 1 / 5 \\
ChatGPT & 4 / 3 & 0 / 3 \\
CV-only & 2 / 2 & 2 / 1 \\
\bottomrule
\end{tabular}
\caption{\label{table:topic-distribution}
Topic Distribution in top-40 Papers. The table shows the number of papers for each topic in the top-20 and 21-40 ranked papers for the \june{} and \sept{} datasets.}
\end{table}
\zrf{Table \ref{table:topic-distribution} summarizes the topic distribution of the top-40 papers from both \june{} and \sept{} datasets. }
\begin{itemize}
\item Of the top-20 papers, 30\% are \textbf{technical reports on LLMs}, including LLaMA models (\cite{touvron2023llama}, \cite{touvron2023llama2}), PaLM 2 \cite{anil2023palm}, PaLM-E \cite{driess2023palme}, and GPT-4 \cite{openai2023gpt4}. GPT-4 has become the paper with the highest z-score, surpassing LLaMA which led the previous ranking. However, Llama 2,
released more recently in mid-July, has seen a rapid increase in citations, reflecting an overall dominance of LLaMA models. Moreover, since the last report, the rankings have been updated to also include MiniGPT-4 \cite{zhu2023minigpt}, a vision-language model that integrates components from Vicuna \cite{chiang2023vicuna}, which is built upon \se{the} LLaMA architecture. 
More generally, the rest of the top-40 list has also witnessed a rise in the number of technical reports, with new models like \textit{Baichuan 2} and \textit{Otter}. 

\item The majority of the papers is focused on \textbf{LLMs}, with 90\% of the top-40 papers (36 out of 40) dedicated to this area. 
\item Notably, five 
out of the top-20 papers discuss multimodality \se{while}\se{,} as shown in Table \ref{table:top20june}, the previous top-20 list contains three. 
\zrf{We should keep} in mind our expanded search criteria to also include especially cs.CV. For example, \se{models} like BLIP-2 \cite{li2023blip} have entered the ranking, 
\se{which proposes} more compute-efficient approaches to synergize vision and LLMs. 

\item The focus on \textbf{problem-solving} within the domain of LLMs is represented by two papers in the top-20 list: \textit{Tree of Thoughts} \cite{yao2023tree}, which continues to maintain its high position in the rankings; and \textit{Large Language Models as Optimizers} \cite{yang2023large}, submitted in early September, which explores the use of LLMs in optimization tasks. 

\item \textbf{Benchmarking and evaluation} of LLMs also remain a strong focus in the field. Two papers from the top-20 list are dedicated to this topic: \textit{Judging LLM-as-a-Judge with MT-Bench and Chatbot Arena} \cite{zheng2023judging}, which has improved by 6 ranks since the last report, and a newly introduced paper \textit{Lost in the Middle} \cite{liu2023lost}, which \zrf{evaluates} LLM performance with lengthy inputs. 
\textit{Universal and Transferable Adversarial Attacks on Aligned Language Models}, \cite{zou2023universal}, newly introduced in this ranking, delves into the evaluation from \textbf{security and ethical dimensions}, examining adversarial strategies to enhance model defenses against producing detrimental content.

\item \textbf{Survey papers} included in our report concentrate on varying aspects of LLMs, such as their limitations (\cite{zhang2023siren}), evaluation (\cite{chang2023survey}), and applications (\cite{xi2023rise}). Meanwhile, we also observe a substantial decline in the rankings of some established surveys (e.g.\ \textit{A Survey of Large Language Models} \cite{zhao2023survey}, submitted in late March). Surveys such as \textit{Harnessing the Power of LLMs in Practice: A Survey on ChatGPT and Beyond} and \textit{Augmented Language Models: a Survey}) published earlier \se{in 2023} have left the top-40 list. Interestingly, both augmented LLMs and toolformer share similar topics. \zrf{This may suggest a decline of interest in research concerning augmented LLMs.} 

\item A shift away from research centered exclusively on \textbf{ChatGPT} (a trend which has been already seen in the previous report)
has become even more apparent in the last 3 months.
More specifically, in the current dataset, there are no papers specifically focused on ChatGPT within the top-20, and only three make it to the top-40 (ranks 26, 33, and 38). This change point\se{s} towards a diversification of interest within LLM research, moving beyond a singular model. 

\end{itemize}

\se{Our} analysis indicates that foundation models like GPT-4 and LLaMA remain influential, coupled with a broader research scope that includes multimodality, specialized LLMs, their application and evaluation.

\subsection{\se{Further} \se{a}nalysis}

In this section, we expand on the previous findings of a marked focus on LLMs, multimodality, and related topics in the top-40 papers and explore the evolution of these trends through a detailed analysis of keyword prevalence. Furthermore, we 
assess institutional contributions, as well as the \zr{global} distribution of research \zr{institutions} in the field.

\subsubsection{How popular are LLMs and multimodality?} 
\begin{figure}[!tb]
    \centering
    \includegraphics[width=\textwidth]{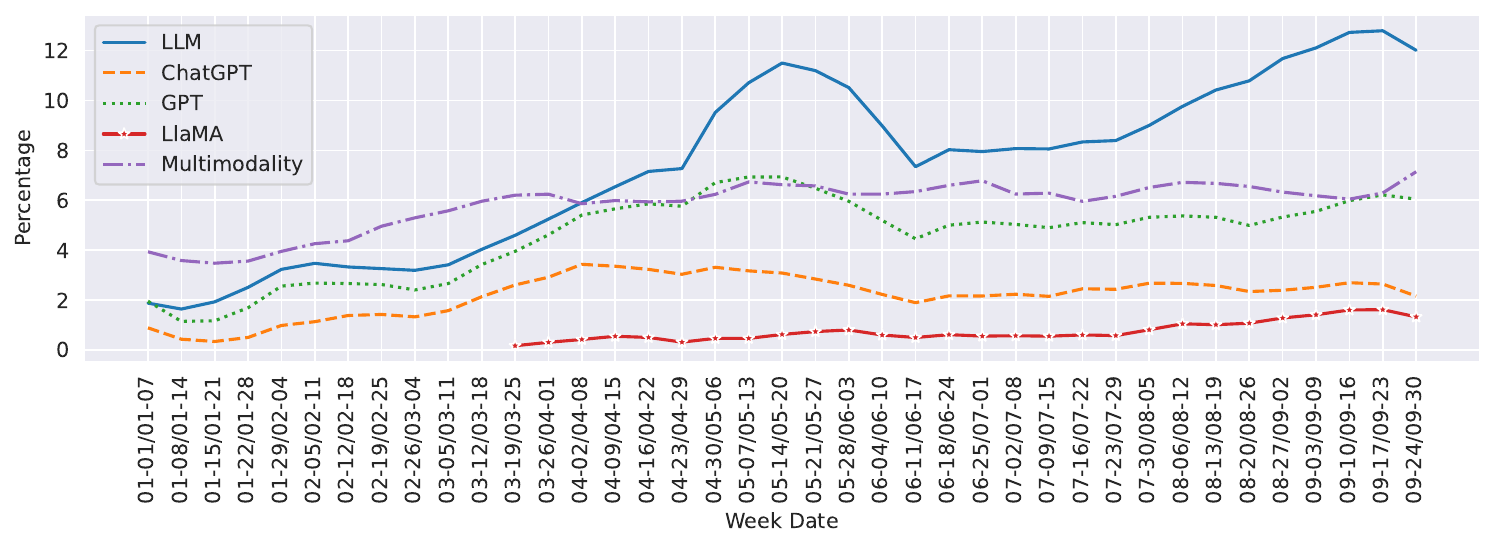}
    \caption{Popularity of ChatGPT, GPT, LLMs and multimodality (in the percentage of papers containing the words in their abstracts or titles) over time in our dataset. We smooth the curve with Savitzky-Golay filter with a window length of 8 and polynomials of order 3.}
    \label{fig:popularity}
\end{figure}

We delve into the developing trend of keywords discussed in the previous section. Figure \ref{fig:popularity} shows how the popularity of each keyword has developed over time in our complete arXiv dataset \arxivlarge{}. We query the keywords ``LLM'', ``ChatGPT'', ``GPT'', ``LLaMA'' and ``Multimodality'' in our dataset over time and flag the paper as relevant if it contains the keywords in its title or abstract.\footnote{We lowercase both abstracts and titles and look for the keywords ``llm(s)'' and ``large language model(s)'' for LLM; ``chatgpt'' and ``chat-gpt'' for ChatGPT; ``gpt'' for GPT; ``llama'' for LLaMA and ``multimodal(ity)'', ``multi-modal(ity)'', ``text-to-image'', ``visual-language'', ``captioning'', ``image-to-text'' for Multimodality.} We smooth the curve with Savitzky-Golay filter \cite{press1990savitzky} with a window length of 8 and polynomials of order 3. 


Both ``LLM'', ``GPT'' and ``ChatGPT'' have a comparatively low relevance in January with less than 2\% of papers discussing these topics (LLaMA was first introduced at the end of February) while ``Multimodality'' has a better start with 4-5\% of papers focusing on relevant topics. Their trends over time also differ greatly: 1) (up-stagnate-down) The ChatGPT curve increases until late March (3.5\% of all papers) and stagnates around that level until the last week of May where it peaks at 4.25\% due to the influence of \se{the} popular NLP conference deadline (ACL on the 21st of May). Later it goes down to 2.5\% and stabilizes until the end of September. 2) (up) The curves for LLM, LLaMA, GPT, and multimodality are quite distinct in terms of percentage level. However, all four curves have upward trends where the LLM curve surges from less than 2\% to over 10\% starting mid-August (almost 15\% in late September). The LLM curve surpasses the multimodality curve in mid-April and it peaks in late May \zr{reaching }nearly 17.5\%. Meanwhile, the multimodality curve increases linearly from 4\% to over 7\% at a slower speed compared to LLM. The GPT curve experiences rapid growth following the LLM curve closely until late May and loses momentum with only a slow growth since June. Since its introduction in late February, LLaMA has experienced slow growth, reaching approximately 2\% by September. It has come remarkably close to matching the level of ChatGPT. \zrf{It is worth noting that the GPT curve also contains the keywords for ChatGPT. If we look at both GPT and ChatGPT curves, we observe an upward trend for the GPT starting in June while the ChatGPT curve is stagnating. This may indicate the gradual shift of research focus to other GPT models e.g.\ GPT4 other than ChatGPT.}

\zrf{We also calculate the percentage of the top-40 papers containing the above-mentioned keywords where 67.5\% of the top-40 papers mention LLM,  45\% for GPT, 25\% for ChatGPT and multimodality, and lastly, 15\% for LLaMA. Compared to all papers from \arxivlarge{}, top papers are more concentrated on discussions involving LLM, predominantly, GPT models (45\%) including ChatGPT (25\%) and multimodality (25\%).}

\subsubsection{\se{K}ey trends in institutional \zrf{distribution}} 

\begin{figure}[!htb]
    \centering
    \includegraphics[width=0.75\textwidth]{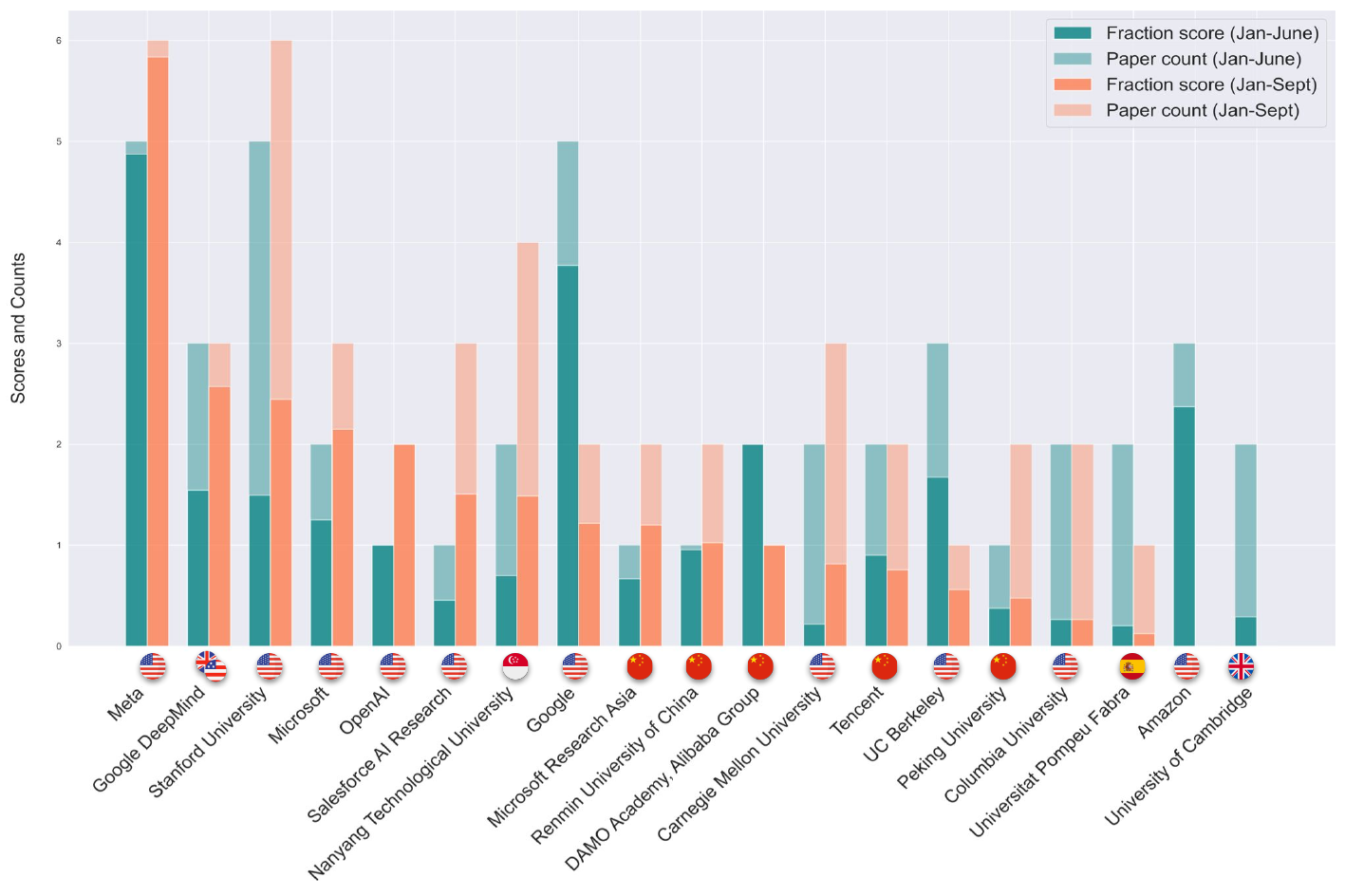}
    \caption{Volume of overall activity (paper count) and their impact (fraction score) by affiliation in top-40 papers: \june{} and \sept{} datasets (sorted by \sept{} fractional count). Google's count includes its Google Research unit. Meta's count includes Meta AI Research and FAIR. Microsoft's count includes Microsoft Research. Tencent's count includes Tencent AI lab, Tencent Cloud AI and ARC Lab, Tencent PCG. \zrf{We include affiliations that have a paper count of at least two in either the \june{} or \sept{} datasets.}}
    \label{fig:big_institutions}
\end{figure}

First, we examine the dynamics of institutions that are the most frequent contributors in the top-40 papers. Secondly, we analyze the country-wise distribution of institutions on a wider scale by industry and academia sectors. Finally, we present a world map showing the global distribution.
 
To compare the presence of institutions in the top-40 list, we plot a bar chart based on the affiliations information indicated in papers from the previous and current datasets (see Figure \ref{fig:big_institutions})\footnote{The figure does not reflect the composition changes within the rankings — such as which papers have remained, disappeared or newly entered the ranking.}. 
This analysis is based on two metrics: the \textbf{cumulative count} of research papers involving a particular institution (indicating its overall activity and presence in the research community), and the \textbf{fractional score} (representing the relative contribution of an institution to individual papers)\se{.}\footnote{The fractional score is computed by dividing an institution's specific contribution by the total number of authors or contributors for a given paper. For instance, in the \textit{Toolformer} paper, 7 authors are affiliated with Meta AI Research and one with Universitat Pompeu Fabra, resulting in fractional scores of \se{$7/8=0.88$ }and \se{$1/8=0.12$}, respectively. \akf{If all authors of a paper are from the same affiliation, then the fractional score for that institution is 1.0. For example, if 24 out of 24 authors are from meta, the fractional score is $24/24=1.0 $ for meta.}}



\paragraph{What are the most frequent contributors to the top-40 list?} 
\begin{itemize}
    \item \textbf{Meta} is leading with six papers and their presence in top research is increasing, evident not only in the introduction of Llama 2 \cite{touvron2023llama2}, but also in their work in the visual domain with DINOv2 \cite{oquab2023dinov2}, a foundational model that generates universal features for image-based tasks. A key distinction \se{of}
    Meta is their tendency to publish papers independently (i.e.\ without collaboration from authors outside the company), suggested by their high fractional score compared to their total paper count. 
    \item \textbf{Stanford University} is also prominent in terms of the number of papers in the top-40 list. However, their contribution is mostly marked by their collaborative projects with other institutions, as seen in papers e.g.\ \textit{Lost in the middle} \cite{liu2023lost} and \textit{Direct Preference Optimization: Your Language Model is Secretly a Reward Model} \cite{rafailov2023direct}. The only independent publication from Stanford, \cite{zhang2023adding}, focuses on text-to-image diffusion models, and this paper is newly introduced to the top-40 list.
    \item \textbf{DeepMind}'s influence is highlighted by their continued input following the April 2023 merger with Google Brain: among their recent standalone works, \textit{Optimization by PROmpting (OPRO)} \cite{yang2023large} (an innovative method leveraging LLMs for optimization tasks framed in natural language) makes it to the top-20 list shortly after the publication in early September. Conversely, \textbf{Google}'s representation has shrunk (both by fraction score, as well as by total paper counts), with only \textit{PaLM 2} \cite{anil2023palm} and \textit{PaLM-E} \cite{driess2023palme} remaining in the top-40, while earlier papers released in January, including works on FLAN models, have left the top ranking. 
    \item \textbf{UC Berkeley} has also seen a modest reduction in its presence among the top-40 papers. However, it is worth noting that the paper \textit{Judging LLM-as-a-Judge with MT-Bench and Chatbot Arena} (which is first authored by UC Berkeley) moved up to rank 5.
    \item \textbf{OpenAI} have maintained their presence in the top-40 list with an increase of one paper authored exclusively by the company since the last report \cite{lightman2023let}. \textbf{Microsoft}, along with \textbf{Microsoft Research Asia}, have maintained their presence, especially in the papers centered around LLMs. 
    \item \textbf{Nanyang Technological University} stands out for its high volume of research papers. Although the ratio between their total number of papers and fractional score (1.49 score for 4 papers) suggests that they frequently collaborate with other institutions (mostly with universities), their independent research impact has recently grown, with the paper \textit{Otter: A Multi-Modal Model with In-Context Instruction Tuning} entering the top ranking. Similarly, \textbf{Carnegie Mellon University}, while also having a high total paper count, usually contributes through co-authorship, predominantly with universities. 
\end{itemize}

In the current dataset, it is notable that there are \textbf{no new affiliations} entering the \sept{} top-40 list. At the same time, two institutions have disappeared from the list: 1) Amazon, previously contributing papers out of independent research such as \textit{Are Emergent Abilities of Large Language Models a Mirage?} and \textit{SemEval-2023 Task 2: Fine-grained Multilingual Named Entity Recognition (MultiCoNER 2)}; 2) University of Cambridge, which often collaborates with various universities and research institutes.

\paragraph{What is the distribution of academic vs.\ industrial representation across regions in the top-40 list \zrf{using fractional score}?} 
As illustrated in Figure \ref{fig:big_institutions}, 12 out of 19 institutions are US-based 
and half of them are companies. Among institutions with a fractional score above 2 for both analyzing periods, all 6 institutions are companies except for Standford University. Following the US, Chinese institutions mark their presence, characterized by a mix of independent research and collaborative efforts.

To explore this pattern further, Figure \ref{fig:big_countries_institutions} showcases the geographical distribution and highlights the industrial and academic impact of all institutions in the top-40 list.\footnote{In our analysis, ``academia'' encompasses universities and research entities, e.g.\ CZ Biohub. It is also important to note that certain less common contributors, such as independent authors or a single case of a governmental organization are not included in this industry-academia analysis by geographical areas (countries or regions).} Our findings include: 
\begin{itemize}
\item \textbf{\textit{Finding 1: The US dominates in both industry and academia followed by China.}}\\
The United States exhibits a pronounced dominance in the top list in both industry and academia. American companies have achieved a high fractional score of 16.10, with a total of 20 papers in \june{} top-40 list. This has slightly adjusted in the \sept{} top-40 list to a score of 14.54 and 19 
out of 40 papers. 
Similarly, American institutions in academia initially held a score of 8.42 across 13 papers, which has increased in the current report to 9.98 for 14 papers. Both pieces of evidence point to the continued or even strengthened dominance of the US in research. Chinese institutions come closest to the second place and demonstrate a balanced involvement in industry and academia with a fractional score of 4.69 over 7 papers for companies, and 4.94 across 8 papers for academia.  
These efforts often involve partnerships with international (predominantly American) institutions. A notable instance of such collaboration is the joint work on \textit{A Survey on Evaluation of Large Language Models} in rank 29, involving several Chinese universities like Jilin University, Westlake University, and Hong Kong University of Science and Technology, alongside American institutions such as Carnegie Mellon University and the University of Illinois at Chicago. 

In contrast, Europe's representation is relatively limited, both in terms of total paper count and fractional score. The UK stands out. This is primarily due to contributions from DeepMind, which later merged with Google Brain (the US company); Germany ranks second after the UK with academia represented by universities such as Leipzig University, TU Munich, TU Berlin and companies such as Bosch Center for AI and SAP; Spain also shows some presence, led by Universidad Pompeu Fabra (academia), and a minor
contribution from Telefonica I+D from industry in the Starcoder paper \zrf{(one out of 67 authors)}. European representation also includes Sweden (one industry institution), Switzerland (one industry and one academia institution), France (one academia), and Austria (one academia). This is further reflected in their fractional scores, which are low due to frequent co-authorship in papers like Starcoder. 

\begin{figure}[!htb]
    \centering
    \includegraphics[width=0.75\textwidth]{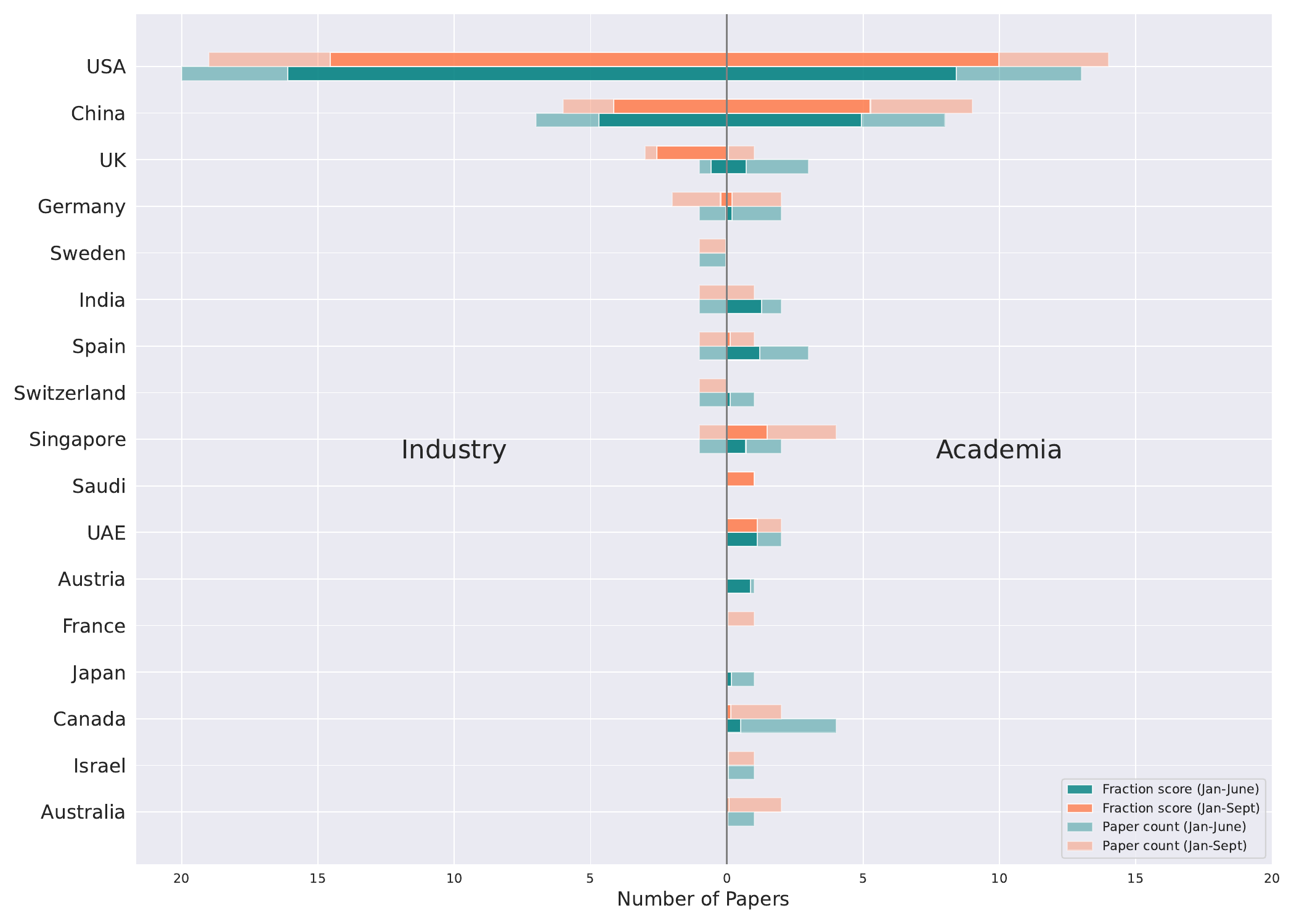}
    \caption{Country-wise distribution of the top-40 papers by institutional sector, showing the number of papers and fractional scores for \june{} and \sept{} datasets}
    \label{fig:big_countries_institutions}
\end{figure}
\item \textbf{\textit{Finding 2: Academia shows a more diverse country-wise presence than industry.}}
Countries other than the US and China primarily demonstrate their presence in academic research, with limited industrial contributions. An exception is again the UK, which has shown an increased industrial contribution in the \arxivlarge{} dataset. Interestingly, this is largely due to the contribution from DeepMind, which was mentioned earlier. The industrial contribution of other countries often involves just one or two authors participating in larger projects, which is also reflected in the relatively low fractional scores for their companies. For instance, the first author of \textit{PaLM-E} is affiliated with both Robotics at Google and TU Berlin. Similarly, Universitat Pompeu Fabra contributes one author to \textit{Toolformer} alongside Meta AI Research. There is notable diversity and dynamism among countries outside the US and China. For instance, while some countries like Singapore are seeing a substantial rise in both score and paper count, other countries like Israel and Australia also present modest academic outputs or newly enter the top-40 list (France and Saudi Arabia).

\item \textbf{\textit{Finding 3: Industry has a substantial presence in the top-40 paper.}}
\begin{table}[!htb]
    \begin{subtable}{.5\linewidth}
      \centering
\scriptsize
\begin{tabular}{lccc}\toprule
&\textbf{Industry} & \textbf{Academia} \\\midrule
\textbf{Independent Research }&14 &6 \\
&\multicolumn{2}{c}{\textit{Collaboration}} \\
\textbf{-with Industry} &0 & 15 \\
\textbf{-with Academia} &15 &5 \\
\bottomrule
\end{tabular}
\caption{}
\label{tab:indu_aca}
    \end{subtable}%
    \begin{subtable}{.5\linewidth}
      \centering
        \centering
\scriptsize
\begin{tabular}{lccc}\toprule
&\textbf{Industry} & \textbf{Academia} \\\midrule
\textbf{Independent Research }&18.6 & 9.3 \\
&\multicolumn{2}{c}{\textit{Collaboration}} \\
\textbf{-with Industry} &0 & 16.9 \\
\textbf{-with Academia} &16.9 &14.6 \\
\bottomrule
\end{tabular}
\caption{}
\label{tab:author_avg}
    \end{subtable} 
    \caption{(a) Count of independent or collaborative contributions of the top-40 papers. (b) The average number of authors by each type of contribution in the top-40 papers.}
\end{table}
Table \ref{tab:indu_aca} shows the count of independent or collaborative contributions of the top-40 papers. Our results indicate that 50\% out of 40 papers are independent research authored by researchers from one institution. Among them, \textbf{70\%} (14 out of 20) are from industry (companies), e.g.\ \textit{LLM as Optimizers} from Google Deepmind and many technical reports such as \textit{GPT-4 Technical Report} from OpenAI, \textit{LLaMA} from Meta and \textit{Baichuan} from Baichuan Inc. 6 independent works come from academia, e.g.\ \textit{MiniGPT-4} from King Abdullah University of Science and Technology, \textit{QLoRA} from the University of Washington and \textit{The Rise and Potential of Large Language Model Based Agents: A Survey} from \se{the} Fudan NLP Group. Interestingly, the industry sector never collaborates with each other in our top-40 papers. This is \zrf{expected} due 
to their competitive positions in business. Academia are more welcoming in terms of cooperation. Among the 20 collaborative papers, 5 (25\%) are collaborations between academic institutions while \textbf{75\%} are collaborating with the industry. Overall, we observe 29 out of 40 papers (nearly 75\%) with industry involvement. 

It is also worth noting that papers from industry or with institutional collaboration, predominantly industry, tend to have a larger number of co-authors as shown in Table \ref{tab:author_avg}. Independent research from industry (i.e.\ without collaboration with other institutions) have the highest average number of authors \se{(}18.6\se{)}, which is 
twice
the average author counts of independent research from academia (9.3). Collaborative works between industry and academia have an average of 16.9 authors while collaborative papers within academia have a slightly lower count of 14.6 on average. In general, the top-40 papers have an average of 16.05 authors, with a notably high standard deviation of 16.21, while the remaining papers have a more modest average of 4.72 authors, with a standard deviation of 3.70. This could be due to the large proportion of industry involvement and collaboration in the top-40 list. 
Interestingly, the number of authors on average from the \sept{} top-40 list has also increased compared to the \june{} top-40 list with an average of 11.8 authors and a standard deviation of 19.5. 
\end{itemize}

\begin{figure}[!ht]
    \centering
    \includegraphics[width=0.95\textwidth]{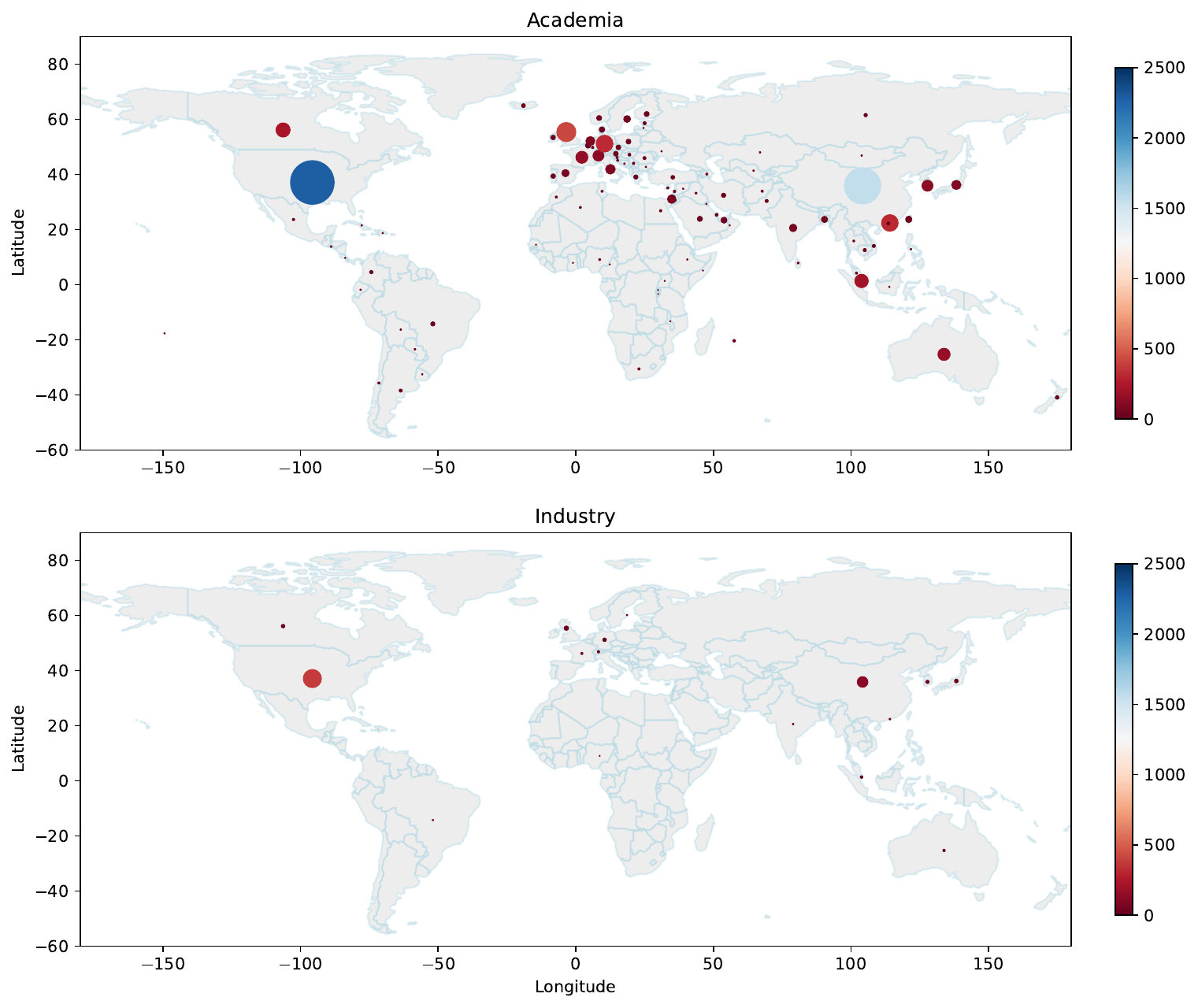}
    \caption{Global distribution of the top-9k papers by institutional sector, showing the proportional counts (presented by bubble colors). The size of the bubbles is proportional to the count. For illustration purposes, we enlarge the bubble sizes for the industry sector.}
    \label{fig:world_map}
\end{figure}

\paragraph{What is the global distribution of academic and industrial research contribution and how does it compare to the top-40 \zrf{using proportional count?}} 
We zoom out from the top-40 papers and analyze our findings in a wider context. We extract institutional information from the top-9k papers using \se{the} GROBI\ak{D} \cite{lopez2009grobid} CRF Wapiti model which has an accuracy rate of 85.8\% for affiliation-address information extraction according to their documentation.\footnote{\url{https://grobid.readthedocs.io/en/latest/Benchmarking-models/\#accuracy}}
For every paper, we identify all institutions involved and count them only once regardless of the number of authors associated with it. 
We exclude papers that contain no affiliation extraction due to file corruption and papers with unidentifiable institutions (e.g.\ affiliations using abbreviations without identifiable addresses). Our final dataset contains 7478 papers with 2293 institutions. We classify these institutions according to their geographical information and sectors (academia or industry). Figure \ref{fig:world_map} shows the distribution of proportional
count for institutions by sectors across their geographical locations.\footnote{We give each institution a \textbf{proportional 
count} based on institution uniqueness. For example, the Toolformer paper contains two affiliations: Meta AI and Universitat Pompeu Fabra. We would assign 0.5 to each affiliation. Note that the proportional count is different from the \textbf{fractional score} we assign to in \ref{fig:big_countries_institutions}.} 

Overall, \textit{academia shows a more diverse \zrf{global} presence than the industry.} This is evident if we compare the industry-focused world map with the academia-focused one: the latter displays a diverse spread of bubbles across the globe indicating various degrees of involvement of academic institutions. In contrast, the world map representing the industrial sector shows a much sparser distribution with only several observable bubbles from Europe, North America, East Asia, and Central Asia. Furthermore, the global distribution involves contributions from 101 countries, whereas the top-40 papers \se{involves} just 17 countries. 

\textbf{\textit{Global distribution by continent.}}
\begin{itemize}
    \item \textbf{North America}: For both academia and industry, there is a pronounced concentration in the US. For example, in academia, Canada's contribution is represented by a proportional count of 219.8, which is only one-tenth of that in the US (2283.4). In contrast, Canada's industrial presence is notably smaller, accounting for just 9.9 proportional counts, which is roughly one-thirtieth of that of the US (374.0).
    \item \textbf{Europe}: Both maps show notable activity in Europe, with multiple bubbles spread across the continent. This suggests a robust presence of both academic institutions and industries in this area. For academia, there are several medium to large clusters located in Central and Western Europe. For example, Germany holds a proportional count of 318.1 in academia with more than 150 academic institutions contributing to the list e.g.\ TU Munich (40.1), University of Tübingen (22.4, predominantly Tübingen AI Center), LMU (22.3), RWTH Aachen (16.7) and University of Stuttgart (16.1) while the proportional count for the German industry is notably lower, standing at just 9.5. Major contributors 
    to this count include Robert Bosch GmbH (3.7), Siemens AG (2.7), SAP (1.0), and the Mercedes-Benz Group (0.7), among others. France, Switzerland, and Italy also demonstrate a strong \textbf{academic presence}, with proportional counts of 155.4, 124.6, and 96.4, respectively. 
    \item \textbf{Asia}: A clear presence in \textbf{East Asia} is also observed in both industry and academia, with institutions mainly originating from \zrf{Mainland China (1561.1 in academia and 120.5 in industry), Hong Kong (309.8 in academia and 1.0 in industry), Japan (88.2 in academia and 10.7 in industry), and South Korea (128.6 in academia and 7.7 in industry)}. Singapore, located in \textbf{South Asia}, stands out as a strong competitor in the academia sector with a proportional count of 197.7. \textbf{Central Asia} academia is led by the United Arab Emirates with a proportional count of 32.7 contributed by research institutions such as Mohamed bin Zayed University of Artificial Intelligence and New York University Abu Dhabi. 
    \item In other regions, such as Australia, India, Brazil and \textbf{South Africa}, we observe noticeable but smaller clusters in both academia and industry. This suggests that while these areas have an influential presence, it is not as concentrated as in the previously mentioned regions. Large areas of \textbf{Africa}, \textbf{Central Asia}, and parts of \textbf{South America} show little to no activity in both sectors. 
\end{itemize}


\begin{table}[!htb]
    \begin{subtable}{1\linewidth}
      \centering
\scriptsize
\begin{tabular}{lrrrrrr}
\toprule
\textbf{Sector}&\textbf{Sector Total} &\textbf{US} &\textbf{CN} &\textbf{Europe} &\textbf{The rest} \\\midrule
\textbf{Academia} &54.63\% &22.93\% &15.85\% &3.81\% &12.04\% \\
\textbf{Industry} &45.37\% &31.56\% &6.38\% &1.10\% &6.34\% \\
\textbf{Region \zrf{Total}}  & 100.00\% &54.49\%	& 22.23\% &	4.91\% & 18.37\% \\
\bottomrule
\end{tabular}
\caption{Top-40 papers.}
\label{tab:pptop40}
    \end{subtable}%
    \bigskip
    \\
    \begin{subtable}{1\linewidth}
      \centering
        \centering
\scriptsize
\begin{tabular}{lrrrrrr}
\toprule
\textbf{Sector} &\textbf{Sector Total} &\textbf{US} &\textbf{CN} &\textbf{Europe} &\textbf{The rest} \\\midrule
\textbf{Academia} &92.46\% &30.61\% &25.12\% &18.38\% &18.35\% \\
\textbf{Industry} &7.54\% &5.02\% &1.63\% &	0.36\% & 0.54\% \\
\textbf{Region \zrf{Total}} &100.00\% &35.63\% &26.75\% &18.74\% &	18.89\% \\
\bottomrule
\end{tabular}
\caption{Top-9k papers}
\label{tab:pptop9k}
    \end{subtable} 
    \caption{(a) The percentage contribution of institutional sectors and regions based on proportional count for the \sept{} top-40 papers. (b) The percentage contribution of institutional sectors and regions based on proportional count for  the \sept{} top-9k papers.}
    \label{tab:ppcount}
\end{table}

\paragraph{\textit{Global distribution vs.\ top-40 distribution}}

Observations from the world map support our earlier discussion about the most frequent contributors to the top-40 papers in the last paragraphs. For a comparable result between the global distribution, i.e.\ top-9k papers\se{,} and top-40 papers, we calculate the percentage contribution of both datasets based on \textit{proportional counts} by key regions (the US, China, European countries and the rest of the world) for industry and academia sectors. The results are shown in Table \ref{tab:ppcount}.

We observe again that \textit{the US \se{clearly dominates} among the top-9k papers, followed by China in both industry and academia.} Table \ref{tab:pptop9k} illustrates this point, where the US is leading with 35.63\% of all proportional counts of the top-9k papers. Sector-wise, for academia, the US contributes 33.1\% ($=30.61\%/92.46\%$) of all proportional counts in 
\se{the} academia sector. Particularly in industry, the US takes up 66.6\% ($=5.02\%/7.53\%$) of the total industry contribution in top-9k papers. China takes up the second place. While Europe as a whole ranks third in academia, their industry representations are much lower both in rank and in percentage. \zrf{It is also worth noting that the ratio of the region total percentage for the US (35.65\%) to that of Europe (18.74\%) is 2:1 in top-9k papers while in top-40 papers, this ratio is drastically increased to over \textbf{10:1} (54.49\%:4.91\%). This is due to both the insufficient presence of Europe academia compared to other regions and the outstanding performance of the US industry in the top-40 papers.} 

Table \ref{tab:ppcount} also confirms \textbf{the substantial presence of the industry in the top-40 papers compared to the top-9k papers}. \zrf{In the top-9k papers, academia dominates with 92.46\% of all proportional counts, compared to only 7.54\% for industry. However, in the top-40 papers, academia contributes 54.63\% of all proportional counts with industry at a level of 45.35\%.} This suggests that compared to the global picture, the top-40 papers show a stronger skewness towards industry, led by the US.

\zrf{To conclude the analysis of} the institutional distribution within the top-40 papers and the top-9k papers of the dataset \arxivlarge{}, we observe that: i) the US clearly dominates in both top-40 and top-9k papers; ii) \zrf{Europe's presence in academia is severely lacking in top-40 papers compared to that of the top-9k papers;} iii) industry representation is largely centered around the US, evident from both the top-40 and global analysis; iv) the global trends indicate more widespread activity in academia compared to the industry.

\section{Conclusion}\label{sec:conclusion}

We have examined arXiv papers related to the categories cs.CL, cs.LG, cs.AI and cs.CV over the first 9 months of 2023. We sorted the papers according to their normalized citation counts with score stabilization, \se{finding} \ak{continued dominance of LLM-related papers}. Our analysis encompassed: (i) dataset statistics, citation distributions over time, and the characteristics of top papers compared to those outside the top-40 list; (ii) an in-depth examination of the top-40 papers, revealing 
shifts in trends and rankings compared to our previous report with a continued dominance of LLMs in NLP and an increasing focus on multimodality; (iii) the popularity of keywords including foundation models such as ChatGPT (GPT) \& LLaMA, LLM and multimodality (these topics have either `caused' the hype surrounding LLMs in late 2022 or have emerged as key subjects of recent discussions); (iv) trends in institutional contribution, which reveal US dominance, especially in industry, whereas academic contribution is more geographically diverse. 

Given the growing popularity of AI and its subfields \cite{ziems2023can}, we hope that our investigation is beneficial not only to newcomers, delivering quick links to useful starting literature and directions, but also to established researchers and their doctoral students. In the future, we plan to regularly update the current report to track shifts of research focus over time and to examine our arXiv datasets \arxivlarge{} and \arxivsmall{} in more depth.

\section*{Acknowledgements}
We thank Adia Khalid, Alina Deriyeva and Jesper Dannath from AG Knowledge Representation and Machine Learning for their valuable review and feedback on this report. We \se{further} thank Sotaro Takeshita and Tornike Tsereteli from DWS, University of Mannheim for their thoughtful discussion. The NLLG group gratefully acknowledges support from the Federal Ministry of Education and Research (BMBF) via the interdisciplinary AI research grant ``Metrics4NLG''. Steffen Eger is further supported by the DFG Heisenberg grant EG 375/5-1.
\bibliographystyle{plain}
\bibliography{my} 

\begin{thebibliography}{10}

\bibitem{aksnes2019citations}
Dag~W Aksnes, Liv Langfeldt, and Paul Wouters.
\newblock Citations, citation indicators, and research quality: An overview of basic concepts and theories.
\newblock {\em Sage Open}, 9(1):2158244019829575, 2019.

\bibitem{anil2023palm}
Rohan Anil, Andrew~M Dai, Orhan Firat, Melvin Johnson, Dmitry Lepikhin, Alexandre Passos, Siamak Shakeri, Emanuel Taropa, Paige Bailey, Zhifeng Chen, et~al.
\newblock Palm 2 technical report.
\newblock {\em arXiv preprint arXiv:2305.10403}, 2023.

\bibitem{bang2023multitask}
Yejin Bang, Samuel Cahyawijaya, Nayeon Lee, Wenliang Dai, Dan Su, Bryan Wilie, Holy Lovenia, Ziwei Ji, Tiezheng Yu, Willy Chung, Quyet~V. Do, Yan Xu, and Pascale Fung.
\newblock A multitask, multilingual, multimodal evaluation of chatgpt on reasoning, hallucination, and interactivity, 2023.

\bibitem{brohan2023rt}
Anthony Brohan, Noah Brown, Justice Carbajal, Yevgen Chebotar, Xi~Chen, Krzysztof Choromanski, Tianli Ding, Danny Driess, Avinava Dubey, Chelsea Finn, et~al.
\newblock Rt-2: Vision-language-action models transfer web knowledge to robotic control.
\newblock {\em arXiv preprint arXiv:2307.15818}, 2023.

\bibitem{bubeck2023sparks}
S{\'e}bastien Bubeck, Varun Chandrasekaran, Ronen Eldan, Johannes Gehrke, Eric Horvitz, Ece Kamar, Peter Lee, Yin~Tat Lee, Yuanzhi Li, Scott Lundberg, et~al.
\newblock Sparks of artificial general intelligence: Early experiments with gpt-4.
\newblock {\em arXiv preprint arXiv:2303.12712}, 2023.

\bibitem{chang2023survey}
Yupeng Chang, Xu~Wang, Jindong Wang, Yuan Wu, Kaijie Zhu, Hao Chen, Linyi Yang, Xiaoyuan Yi, Cunxiang Wang, Yidong Wang, et~al.
\newblock A survey on evaluation of large language models.
\newblock {\em arXiv preprint arXiv:2307.03109}, 2023.

\bibitem{chiang2023vicuna}
Wei-Lin Chiang, Zhuohan Li, Zi~Lin, Ying Sheng, Zhanghao Wu, Hao Zhang, Lianmin Zheng, Siyuan Zhuang, Yonghao Zhuang, Joseph~E Gonzalez, et~al.
\newblock Vicuna: An open-source chatbot impressing gpt-4 with 90\%* chatgpt quality.
\newblock {\em See https://vicuna. lmsys. org (accessed 14 April 2023)}, 2023.

\bibitem{douglas2023large}
Michael~R Douglas.
\newblock Large language models.
\newblock {\em arXiv preprint arXiv:2307.05782}, 2023.

\bibitem{driess2023palme}
Danny Driess, Fei Xia, Mehdi~SM Sajjadi, Corey Lynch, Aakanksha Chowdhery, Brian Ichter, Ayzaan Wahid, Jonathan Tompson, Quan Vuong, Tianhe Yu, et~al.
\newblock Palm-e: An embodied multimodal language model.
\newblock {\em arXiv preprint arXiv:2303.03378}, 2023.

\bibitem{eger2023nllg}
Steffen Eger, Christoph Leiter, Jonas Belouadi, Ran Zhang, Aida Kostikova, Daniil Larionov, Yanran Chen, and Vivian Fresen.
\newblock Nllg quarterly arxiv report 06/23: What are the most influential current ai papers?
\newblock {\em arXiv preprint arXiv:2308.04889}, 2023.

\bibitem{girdhar2023imagebind}
Rohit Girdhar, Alaaeldin El-Nouby, Zhuang Liu, Mannat Singh, Kalyan~Vasudev Alwala, Armand Joulin, and Ishan Misra.
\newblock Imagebind: One embedding space to bind them all.
\newblock In {\em Proceedings of the IEEE/CVF Conference on Computer Vision and Pattern Recognition}, pages 15180--15190, 2023.

\bibitem{jakobsen2014thresholds}
Janus~Christian Jakobsen, J{\o}rn Wetterslev, Per Winkel, Theis Lange, and Christian Gluud.
\newblock Thresholds for statistical and clinical significance in systematic reviews with meta-analytic methods.
\newblock {\em BMC medical research methodology}, 14(1):1--13, 2014.

\bibitem{krause1986social}
Neal Krause.
\newblock Social support, stress, and well-being among older adults.
\newblock {\em Journal of gerontology}, 41(4):512--519, 1986.

\bibitem{li2023otter}
Bo~Li, Yuanhan Zhang, Liangyu Chen, Jinghao Wang, Jingkang Yang, and Ziwei Liu.
\newblock Otter: A multi-modal model with in-context instruction tuning.
\newblock {\em arXiv preprint arXiv:2305.03726}, 2023.

\bibitem{li2023blip}
Junnan Li, Dongxu Li, Silvio Savarese, and Steven Hoi.
\newblock Blip-2: Bootstrapping language-image pre-training with frozen image encoders and large language models.
\newblock {\em arXiv preprint arXiv:2301.12597}, 2023.

\bibitem{li2023starcoder}
Raymond Li, Loubna~Ben Allal, Yangtian Zi, Niklas Muennighoff, Denis Kocetkov, Chenghao Mou, Marc Marone, Christopher Akiki, Jia Li, Jenny Chim, et~al.
\newblock Starcoder: may the source be with you!
\newblock {\em arXiv preprint arXiv:2305.06161}, 2023.

\bibitem{lightman2023let}
Hunter Lightman, Vineet Kosaraju, Yura Burda, Harri Edwards, Bowen Baker, Teddy Lee, Jan Leike, John Schulman, Ilya Sutskever, and Karl Cobbe.
\newblock Let's verify step by step.
\newblock {\em arXiv preprint arXiv:2305.20050}, 2023.

\bibitem{liu2023lost}
Nelson~F Liu, Kevin Lin, John Hewitt, Ashwin Paranjape, Michele Bevilacqua, Fabio Petroni, and Percy Liang.
\newblock Lost in the middle: How language models use long contexts.
\newblock {\em arXiv preprint arXiv:2307.03172}, 2023.

\bibitem{lopez2009grobid}
Patrice Lopez.
\newblock Grobid: Combining automatic bibliographic data recognition and term extraction for scholarship publications.
\newblock In {\em Research and Advanced Technology for Digital Libraries: 13th European Conference, ECDL 2009, Corfu, Greece, September 27-October 2, 2009. Proceedings 13}, pages 473--474. Springer, 2009.

\bibitem{newman2014prediction}
Mark~EJ Newman.
\newblock Prediction of highly cited papers.
\newblock {\em Europhysics Letters}, 105(2):28002, 2014.

\bibitem{openai2023gpt4}
OpenAI.
\newblock Gpt-4 technical report, 2023.

\bibitem{oquab2023dinov2}
Maxime Oquab, Timoth{\'e}e Darcet, Th{\'e}o Moutakanni, Huy Vo, Marc Szafraniec, Vasil Khalidov, Pierre Fernandez, Daniel Haziza, Francisco Massa, Alaaeldin El-Nouby, et~al.
\newblock Dinov2: Learning robust visual features without supervision.
\newblock {\em arXiv preprint arXiv:2304.07193}, 2023.

\bibitem{penedo2023refinedweb}
Guilherme Penedo, Quentin Malartic, Daniel Hesslow, Ruxandra Cojocaru, Alessandro Cappelli, Hamza Alobeidli, Baptiste Pannier, Ebtesam Almazrouei, and Julien Launay.
\newblock The refinedweb dataset for falcon llm: outperforming curated corpora with web data, and web data only.
\newblock {\em arXiv preprint arXiv:2306.01116}, 2023.

\bibitem{press1990savitzky}
William~H Press and Saul~A Teukolsky.
\newblock Savitzky-golay smoothing filters.
\newblock {\em Computers in Physics}, 4(6):669--672, 1990.

\bibitem{rafailov2023direct}
Rafael Rafailov, Archit Sharma, Eric Mitchell, Stefano Ermon, Christopher~D Manning, and Chelsea Finn.
\newblock Direct preference optimization: Your language model is secretly a reward model.
\newblock {\em arXiv preprint arXiv:2305.18290}, 2023.

\bibitem{schick2023toolformer}
Timo Schick, Jane Dwivedi-Yu, Roberto Dessì, Roberta Raileanu, Maria Lomeli, Luke Zettlemoyer, Nicola Cancedda, and Thomas Scialom.
\newblock Toolformer: Language models can teach themselves to use tools, 2023.

\bibitem{schmitt1998quantifying}
Norbert Schmitt.
\newblock Quantifying word association responses: What is native-like?
\newblock {\em System}, 26(3):389--401, 1998.

\bibitem{touvron2023llama}
Hugo Touvron, Thibaut Lavril, Gautier Izacard, Xavier Martinet, Marie-Anne Lachaux, Timothée Lacroix, Baptiste Rozière, Naman Goyal, Eric Hambro, Faisal Azhar, Aurelien Rodriguez, Armand Joulin, Edouard Grave, and Guillaume Lample.
\newblock Llama: Open and efficient foundation language models, 2023.

\bibitem{touvron2023llama2}
Hugo Touvron, Louis Martin, Kevin Stone, Peter Albert, Amjad Almahairi, Yasmine Babaei, Nikolay Bashlykov, Soumya Batra, Prajjwal Bhargava, Shruti Bhosale, et~al.
\newblock Llama 2: Open foundation and fine-tuned chat models.
\newblock {\em arXiv preprint arXiv:2307.09288}, 2023.

\bibitem{wu2023visual}
Chenfei Wu, Shengming Yin, Weizhen Qi, Xiaodong Wang, Zecheng Tang, and Nan Duan.
\newblock Visual chatgpt: Talking, drawing and editing with visual foundation models.
\newblock {\em arXiv preprint arXiv:2303.04671}, 2023.

\bibitem{xi2023rise}
Zhiheng Xi, Wenxiang Chen, Xin Guo, Wei He, Yiwen Ding, Boyang Hong, Ming Zhang, Junzhe Wang, Senjie Jin, Enyu Zhou, et~al.
\newblock The rise and potential of large language model based agents: A survey.
\newblock {\em arXiv preprint arXiv:2309.07864}, 2023.

\bibitem{xu2023wizardlm}
Can Xu, Qingfeng Sun, Kai Zheng, Xiubo Geng, Pu~Zhao, Jiazhan Feng, Chongyang Tao, and Daxin Jiang.
\newblock Wizardlm: Empowering large language models to follow complex instructions.
\newblock {\em arXiv preprint arXiv:2304.12244}, 2023.

\bibitem{yang2023baichuan}
Aiyuan Yang, Bin Xiao, Bingning Wang, Borong Zhang, Chao Yin, Chenxu Lv, Da~Pan, Dian Wang, Dong Yan, Fan Yang, et~al.
\newblock Baichuan 2: Open large-scale language models.
\newblock {\em arXiv preprint arXiv:2309.10305}, 2023.

\bibitem{yang2023large}
Chengrun Yang, Xuezhi Wang, Yifeng Lu, Hanxiao Liu, Quoc~V Le, Denny Zhou, and Xinyun Chen.
\newblock Large language models as optimizers.
\newblock {\em arXiv preprint arXiv:2309.03409}, 2023.

\bibitem{yao2023tree}
Shunyu Yao, Dian Yu, Jeffrey Zhao, Izhak Shafran, Thomas~L Griffiths, Yuan Cao, and Karthik Narasimhan.
\newblock Tree of thoughts: Deliberate problem solving with large language models.
\newblock {\em arXiv preprint arXiv:2305.10601}, 2023.

\bibitem{zhang2023adding}
Lvmin Zhang, Anyi Rao, and Maneesh Agrawala.
\newblock Adding conditional control to text-to-image diffusion models.
\newblock In {\em Proceedings of the IEEE/CVF International Conference on Computer Vision}, pages 3836--3847, 2023.

\bibitem{zhang2023siren}
Yue Zhang, Yafu Li, Leyang Cui, Deng Cai, Lemao Liu, Tingchen Fu, Xinting Huang, Enbo Zhao, Yu~Zhang, Yulong Chen, et~al.
\newblock Siren's song in the ai ocean: A survey on hallucination in large language models.
\newblock {\em arXiv preprint arXiv:2309.01219}, 2023.

\bibitem{zhao2023survey}
Wayne~Xin Zhao, Kun Zhou, Junyi Li, Tianyi Tang, Xiaolei Wang, Yupeng Hou, Yingqian Min, Beichen Zhang, Junjie Zhang, Zican Dong, et~al.
\newblock A survey of large language models.
\newblock {\em arXiv preprint arXiv:2303.18223}, 2023.

\bibitem{zheng2023judging}
Lianmin Zheng, Wei-Lin Chiang, Ying Sheng, Siyuan Zhuang, Zhanghao Wu, Yonghao Zhuang, Zi~Lin, Zhuohan Li, Dacheng Li, Eric Xing, et~al.
\newblock Judging llm-as-a-judge with mt-bench and chatbot arena.
\newblock {\em arXiv preprint arXiv:2306.05685}, 2023.

\bibitem{zhou2023comprehensive}
Ce~Zhou, Qian Li, Chen Li, Jun Yu, Yixin Liu, Guangjing Wang, Kai Zhang, Cheng Ji, Qiben Yan, Lifang He, et~al.
\newblock A comprehensive survey on pretrained foundation models: A history from bert to chatgpt.
\newblock {\em arXiv preprint arXiv:2302.09419}, 2023.

\bibitem{zhu2023minigpt}
Deyao Zhu, Jun Chen, Xiaoqian Shen, Xiang Li, and Mohamed Elhoseiny.
\newblock Minigpt-4: Enhancing vision-language understanding with advanced large language models.
\newblock {\em arXiv preprint arXiv:2304.10592}, 2023.

\bibitem{ziems2023can}
Caleb Ziems, William Held, Omar Shaikh, Jiaao Chen, Zhehao Zhang, and Diyi Yang.
\newblock Can large language models transform computational social science?
\newblock {\em arXiv preprint arXiv:2305.03514}, 2023.

\bibitem{zou2023universal}
Andy Zou, Zifan Wang, J~Zico Kolter, and Matt Fredrikson.
\newblock Universal and transferable adversarial attacks on aligned language models.
\newblock {\em arXiv preprint arXiv:2307.15043}, 2023.

\end{thebibliography}

\newpage
\appendix
\section{Appendix}
\subsection{Top-40 lists ranked by stable z-score}
 \begin{table}[!htbp]
 \centering
\fontsize{9.5pt}{9.5pt}\selectfont
 \begin{tabular}{|M{0.5cm}|m{5.5cm}|M{0.9cm}|M{3cm}|M{1.8cm}|M{0.5cm}|M{0.7cm}|}
\hline
\textbf{No.}& \multicolumn{1}{c|}{\textbf{Title}} & \textbf{Cat.}& \textbf{Link} & \textbf{Week} &\textbf{Cit} & \textbf{z-score}  \\ 
\hline
1 &LLaMA: Open and Efficient Foundation Language Models &cs.CL &\url{http://arxiv.org/abs/2302.13971v1} &02-26/03-04 &874 &26.9 \\
2 &GPT-4 Technical Report &cs.CL &\url{http://arxiv.org/abs/2303.08774v3} &03-12/03-18 &509 &24.9 \\
3 &Sparks of Artificial General Intelligence: Early experiments with GPT-4 &cs.CL &\url{http://arxiv.org/abs/2303.12712v5} &03-19/03-25 &354 &23.4 \\
4 &PaLM 2 Technical Report &cs.CL &\url{http://arxiv.org/abs/2305.10403v1} &05-14/05-20 &82 &22.5 \\
5 &PaLM-E: An Embodied Multimodal Language Model &cs.LG &\url{http://arxiv.org/abs/2303.03378v1} &03-05/03-11 &164 &20.2 \\
6 &A Multitask, Multilingual, Multimodal Evaluation of ChatGPT on Reasoning, Hallucination, and Interactivity &cs.CL &\url{http://arxiv.org/abs/2302.04023v2} &02-05/02-11 &214 &16.4 \\
7 &Visual Instruction Tuning &cs.CV &\url{http://arxiv.org/abs/2304.08485v1} &04-16/04-22 &89 &15.3 \\
8 &A Survey of Large Language Models &cs.CL &\url{http://arxiv.org/abs/2303.18223v11} &03-26/04-01 &169 &15.2 \\
9 &QLoRA: Efficient Finetuning of Quantized LLMs &cs.LG &\url{http://arxiv.org/abs/2305.14314v1} &05-21/05-27 &30 &15.1 \\
10 &Segment Anything &cs.CV &\url{http://arxiv.org/abs/2304.02643v1} &04-02/04-08 &165 &15.1 \\
11 &Judging LLM-as-a-judge with MT-Bench and Chatbot Arena &cs.CL &\url{http://arxiv.org/abs/2306.05685v2} &06-04/06-10 &21 &13.8 \\
12 &Voyager: An Open-Ended Embodied Agent with Large Language Models &cs.AI &\url{http://arxiv.org/abs/2305.16291v1} &05-21/05-27 &21 &13.5 \\
13 &Tree of Thoughts: Deliberate Problem Solving with Large Language Models &cs.CL &\url{http://arxiv.org/abs/2305.10601v1} &05-14/05-20 &49 &13.4 \\
14 &How Close is ChatGPT to Human Experts? Comparison Corpus, Evaluation, and Detection &cs.CL &\url{http://arxiv.org/abs/2301.07597v1} &01-15/01-21 &94 &13.3 \\
15 &Extracting Training Data from Diffusion Models &cs.CR &\url{http://arxiv.org/abs/2301.13188v1} &01-29/02-04 &97 &13.1 \\
16 &Toolformer: Language Models Can Teach Themselves to Use Tools &cs.CL &\url{http://arxiv.org/abs/2302.04761v1} &02-05/02-11 &175 &12.3 \\
17 &ImageBind: One Embedding Space To Bind Them All &cs.CV &\url{http://arxiv.org/abs/2305.05665v2} &05-07/05-13 &39 &11.9 \\
18 &HuggingGPT: Solving AI Tasks with ChatGPT and its Friends in Hugging Face &cs.CL &\url{http://arxiv.org/abs/2303.17580v3} &03-26/04-01 &129 &11.2 \\
19 &Is ChatGPT a General-Purpose Natural Language Processing Task Solver? &cs.CL &\url{http://arxiv.org/abs/2302.06476v2} &02-05/02-11 &145 &11.0 \\
20 &A Watermark for Large Language Models &cs.LG &\url{http://arxiv.org/abs/2301.10226v3} &01-22/01-28 &76 &10.8 \\
\hline
 \end{tabular}
 \caption{Papers, their prime category, arXiv link, week of first arXiv submission, citation count (as of 07/27/2023) and \textbf{stable z-score}. \textbf{Top 20 papers} according to \textbf{stable z-score }among all cs.CL and cs.LG papers.}
 \label{table:top20june}
 \end{table}

\begin{table}[!htb]
 \centering
 \footnotesize
 \begin{tabular}{|M{0.5cm}|m{5.5cm}|M{0.9cm}|M{3cm}|M{1.8cm}|M{0.5cm}|M{0.7cm}|}
\hline
\textbf{No.}& \multicolumn{1}{c|}{\textbf{Title}} & \textbf{Cat.}& \textbf{Link} & \textbf{Week} &\textbf{Cit} & \textbf{z-score}  \\ 
 \hline
21 &Mathematical Capabilities of ChatGPT &cs.LG &\url{http://arxiv.org/abs/2301.13867v2} &01-29/02-04 &79 &10.6 \\
22 &Mastering Diverse Domains through World Models &cs.AI &\url{http://arxiv.org/abs/2301.04104v1} &01-08/01-14 &59 &10.5 \\
23 &The Flan Collection: Designing Data and Methods for Effective Instruction Tuning &cs.AI &\url{http://arxiv.org/abs/2301.13688v2} &01-29/02-04 &78 &10.4 \\
24 &DetectGPT: Zero-Shot Machine-Generated Text Detection using Probability Curvature &cs.CL &\url{http://arxiv.org/abs/2301.11305v2} &01-22/01-28 &76 &10.3 \\
25 &The False Promise of Imitating Proprietary LLMs &cs.CL &\url{http://arxiv.org/abs/2303.18223v11} &05-21/05-27 &16 &10.2 \\
26 &Augmented Language Models: a Survey &cs.CL &\url{http://arxiv.org/abs/2302.07842v1} &02-12/02-18 &79 &10.1 \\
27 &Video-LLaMA: An Instruction-tuned Audio-Visual Language Model for Video Understanding &cs.CL &\url{http://arxiv.org/abs/2306.02858v3} &06-04/06-10 &12 &10.1 \\
28 &The RefinedWeb Dataset for Falcon LLM: Outperforming Curated Corpora with Web Data, and Web Data Only &cs.CL &\url{http://arxiv.org/abs/2306.01116v1} &05-28/06-03 &11 &9.8 \\
29 &Large Language Models are not Fair Evaluators &cs.CL &\url{http://arxiv.org/abs/2305.17926v1} &05-28/06-03 &14 &9.8 \\
30 &SemEval-2023 Task 2: Fine-grained Multilingual Named Entity Recognition (MultiCoNER 2) &cs.CL &\url{http://arxiv.org/abs/2305.06586v2} &05-07/05-13 &36 &9.8 \\
31 &PandaGPT: One Model To Instruction-Follow Them All &cs.CL &\url{http://arxiv.org/abs/2301.11305v2} &05-21/05-27 &15 &9.5 \\
32 &mPLUG-Owl: Modularization Empowers Large Language Models with Multimodality &cs.CL &\url{http://arxiv.org/abs/2304.14178v1} &04-23/04-29 &34 &9.0 \\
33 &InstructBLIP: Towards General-purpose Vision-Language Models with Instruction Tuning &cs.CV &\url{http://arxiv.org/abs/2305.06500v2} &05-07/05-13 &33 &9.0 \\
34 &T2I-Adapter: Learning Adapters to Dig out More Controllable Ability for Text-to-Image Diffusion Models &cs.CV &\url{http://arxiv.org/abs/2302.08453v2} &02-12/02-18 &76 &8.8 \\
35 &Employing Multimodal Machine Learning for Stress Detection &cs.LG &\url{http://arxiv.org/abs/2306.09385v1} &06-11/06-17 &21 &8.6 \\
36 &StarCoder: may the source be with you! &cs.CL &\url{http://arxiv.org/abs/2305.06161v1} &05-07/05-13 &28 &8.5 \\
37 &Harnessing the Power of LLMs in Practice: A Survey on ChatGPT and Beyond &cs.CL &\url{http://arxiv.org/abs/2304.13712v2} &04-23/04-29 &32 &8.3 \\
38 &Muse: Text-To-Image Generation via Masked Generative Transformers &cs.CV &\url{http://arxiv.org/abs/2301.00704v1} &01-01/01-07 &111 &8.3 \\
39 &Are Emergent Abilities of Large Language Models a Mirage? &cs.AI &\url{http://arxiv.org/abs/2304.15004v2} &04-23/04-29 &30 &8.2 \\
40 &ChatGPT is not all you need. A State of the Art Review of large Generative AI models &cs.LG &\url{http://arxiv.org/abs/2301.04655v1} &01-08/01-14 &46 &8.2 \\
\hline
 \end{tabular}
 \caption{Papers, their prime category, arXiv link, week of first arXiv submission, citation count (as of 07/27/2023) and \textbf{stable z-score}. \textbf{Top 40 papers} according to \textbf{stable z-score} among all cs.CL and cs.LG papers.}
 \label{table:top40june}
 \end{table}

\subsection{Statistics on arXiv categories}
\emph{Statistics on citations}
\begin{table}[ht]
\fontsize{9.5pt}{9.5pt}\selectfont
    \centering
    \begin{tabular}{c|r}
\hline
\textbf{Category} & \textbf{Occurrences} \\ \hline
cs.CV &14754 \\
cs.LG &13480 \\
cs.CL &8525 \\
cs.AI &2629 \\
eess.IV &1802 \\
stat.ML &1372 \\
cs.RO &1314 \\
cs.CR &753 \\
cs.IR &678 \\
cs.SD &616 \\
cs.HC &540 \\
cs.SE &421 \\
eess.AS &417 \\
eess.SP &416 \\
cs.CY &390 \\
math.OC &370 \\
cs.NE &324 \\
quant-ph &276 \\
eess.SY &272 \\
cs.SI &209 \\
cs.NI &202 \\
cs.DC &190 \\
cs.IT &141 \\
cs.GT &140 \\
stat.ME &140 \\
q-bio.QM &137 \\
q-bio.NC &131 \\
cs.DS &120 \\
math.NA &115 \\
cs.GR &112 \\
cs.DB &111 \\
q-bio.BM &111 \\
\hline
    \end{tabular}
    \caption{All primary categories given in our arXiv dataset whose occurrence exceeds 100. ArXiv categories are described here: \url{https://arxiv.org/category_taxonomy}.}
    \label{table:categories}
\end{table}

\begin{table}[!htbp]
\fontsize{9.5pt}{9.5pt}\selectfont
    \centering
    \begin{tabular}{ccccccccccc} 
    \hline
\textbf{Week Date} & \textbf{Mean}  & \textbf{Std} & \textbf{cs.CL} & \textbf{cs.LG} &\textbf{ cs.CV }&\textbf{ cs.AI} & \textbf{stat.ML} & \textbf{ma|ph }& \textbf{others}\\
\hline
01-01/01-07 &3.66 &13.08 &5.44 &2.27 &6.13 &1.22 &3.06 &1.85 &2.53 \\
01-08/01-14 &3.20 &8.60 &2.84 &3.30 &3.64 &5.85 &1.00 &1.76 &2.79 \\
01-15/01-21 &3.21 &10.34 &7.18 &3.07 &3.48 &2.52 &1.90 &1.81 &1.79 \\
01-22/01-28 &3.33 &9.59 &4.42 &3.23 &3.60 &3.27 &1.78 &3.15 &2.85 \\
01-29/02-04 &4.65 &21.70 &8.94 &2.87 &6.45 &5.08 &2.22 &1.97 &4.91 \\
02-05/02-11 &5.32 &26.79 &14.36 &2.85 &8.31 &3.65 &2.15 &2.30 &2.85 \\
02-12/02-18 &3.55 &10.59 &6.54 &2.60 &4.18 &5.84 &2.34 &1.63 &2.38 \\
02-19/02-25 &3.39 &10.40 &6.64 &2.55 &3.55 &3.41 &1.51 &2.00 &3.28 \\
02-26/03-04 &4.83 &63.44 &23.84 &2.38 &3.13 &2.34 &2.05 &2.73 &2.41 \\
03-05/03-11 &3.32 &14.59 &6.94 &3.12 &3.52 &4.52 &0.87 &2.20 &1.97 \\
03-12/03-18 &4.38 &44.24 &20.33 &2.48 &3.52 &1.55 &1.27 &4.97 &2.12 \\
03-19/03-25 &3.85 &24.03 &13.93 &1.69 &3.79 &4.76 &1.67 &1.19 &1.75 \\
03-26/04-01 &3.95 &19.62 &13.36 &2.44 &2.98 &4.83 &1.61 &2.27 &2.05 \\
04-02/04-08 &3.80 &24.96 &8.24 &1.64 &4.34 &3.08 &1.78 &1.09 &3.21 \\
04-09/04-15 &3.46 &11.15 &10.43 &1.53 &3.02 &4.29 &7.65 &2.86 &2.52 \\
04-16/04-22 &3.35 &14.42 &7.33 &1.77 &3.88 &3.89 &1.44 &1.46 &1.90 \\
04-23/04-29 &2.80 &9.31 &7.30 &1.65 &2.60 &3.09 &1.04 &1.29 &2.14 \\
04-30/05-06 &2.27 &6.20 &3.61 &1.76 &2.64 &1.34 &0.75 &1.11 &1.51 \\
05-07/05-13 &2.42 &9.05 &3.48 &1.16 &3.22 &1.38 &0.67 &1.32 &2.09 \\
05-14/05-20 &2.35 &11.20 &5.00 &1.00 &2.34 &1.41 &0.50 &0.89 &1.40 \\
05-21/05-27 &2.19 &6.74 &2.93 &1.87 &1.65 &4.89 &0.78 &1.02 &1.31 \\
05-28/06-03 &1.66 &5.70 &2.96 &1.32 &1.49 &0.88 &1.00 &0.88 &1.42 \\
06-04/06-10 &1.69 &7.77 &3.37 &0.99 &1.66 &1.50 &0.79 &1.08 &1.45 \\
06-11/06-17 &1.38 &4.11 &2.90 &1.09 &1.39 &0.64 &1.33 &0.58 &0.88 \\
06-18/06-24 &1.21 &4.73 &2.39 &1.11 &1.30 &0.41 &0.52 &0.67 &0.89 \\
06-25/07-01 &1.18 &3.76 &2.44 &0.79 &1.05 &0.93 &0.80 &0.55 &1.18 \\
07-02/07-08 &1.22 &5.04 &2.19 &0.93 &1.12 &0.90 &0.17 &0.58 &1.27 \\
07-09/07-15 &0.95 &3.36 &1.78 &0.49 &0.86 &2.21 &0.28 &1.70 &0.61 \\
07-16/07-22 &1.53 &22.06 &8.87 &0.80 &0.69 &0.80 &0.42 &0.63 &0.74 \\
07-23/07-29 &0.88 &3.96 &1.77 &0.57 &0.68 &2.06 &0.28 &0.54 &0.91 \\
07-30/08-05 &0.85 &3.07 &1.47 &0.70 &0.59 &2.38 &0.40 &0.39 &0.71 \\
08-06/08-12 &1.03 &11.55 &3.74 &0.40 &0.51 &1.13 &1.18 &0.24 &0.69 \\
08-13/08-19 &0.60 &2.04 &2.06 &0.36 &0.46 &0.74 &0.30 &0.42 &0.28 \\
08-20/08-26 &0.80 &7.51 &3.68 &0.32 &0.30 &1.72 &1.07 &0.41 &0.57 \\
08-27/09-02 &0.55 &2.16 &1.16 &0.43 &0.52 &0.47 &0.21 &0.33 &0.48 \\
09-03/09-09 &0.37 &1.63 &0.92 &0.38 &0.22 &0.68 &0.11 &0.24 &0.30 \\
09-10/09-16 &0.40 &1.67 &0.83 &0.40 &0.14 &1.62 &0.13 &0.19 &0.25 \\
09-17/09-23 &0.38 &1.78 &0.81 &0.32 &0.22 &0.30 &0.08 &0.78 &0.28 \\
09-24/09-30 &0.26 &1.64 &0.42 &0.15 &0.21 &0.16 &0.06 &0.25 &0.42 \\
\hline
    \end{tabular}
    \caption{Mean number of citations, over all papers including standard deviations, and for the primary categories cs.CL, cs.LG and the remaining categories.}
    \label{table:mean}
\end{table}

\end{document}